\documentclass[11pt]{article}
\usepackage{natbib}
\usepackage{amsmath,color}
\usepackage{amsfonts}
\usepackage{amssymb}
\usepackage{mathtools}
\usepackage{enumerate}
\usepackage{graphicx,rotating}
\usepackage{subfig}
\usepackage{float}
\usepackage[font=small, margin=1.5cm]{caption}
\title{Adaptive Complementary Ensemble EMD and Energy-Frequency Spectra of Cryptocurrency Prices}
\author{Tim Leung\thanks{Applied Mathematics Department, University of Washington, Seattle WA 98195. Email: {timleung@uw.edu}. Corresponding author.}  \and Theodore Zhao\thanks{Applied Mathematics Department, University of Washington, Seattle WA 98195. Email: {zdzhao16@uw.edu}.}}

\begin{document}
\maketitle
\bigskip

\abstract{We study the price dynamics of cryptocurrencies using adaptive complementary ensemble empirical mode decomposition (ACE-EMD) and Hilbert spectral analysis. This is a multiscale noise-assisted approach that decomposes any time series into a number of intrinsic mode functions, along with the corresponding instantaneous amplitudes and instantaneous frequencies. The decomposition is adaptive to the time-varying volatility of each cryptocurrency price evolution. Different combinations of modes allow us to reconstruct the time series using components of different timescales. We then apply Hilbert spectral analysis to define and compute the   instantaneous energy-frequency spectrum of each cryptocurrency to illustrate the properties of various timescales embedded in the original time series.}

\newpage
\section{Introduction} Empirical   studies and market observations suggest that asset prices are driven by multiscale factors, ranging from long-term market regimes to rapid fluctuations  (see \cite{sircarMultiscaleVolAsy2003,wavelefinancetbook,YAHYA2019277,MAGHYEREH2019895}, among others). Embedded with different timescales, many financial time series often exhibit nonstationary behaviors with trends and time-varying volatilities. These characteristics can hardly be captured by linear models and call for an adaptive and nonlinear approach for analysis. 

One approach  for analyzing nonstationary time series is the Hilbert-Huang transform (HHT) introduced by \cite{Huang1998}. The HHT method can decompose any time series into oscillating components with nonstationary amplitudes and frequencies using the empirical mode decomposition (EMD). This fully adaptive method provides a multiscale decomposition for the original time series, which gives richer information about the time series. The instantaneous frequency and instantaneous amplitude of each component are later extracted using the Hilbert transform. The decomposition onto different timescales also allows for reconstruction up to different resolutions, providing a smoothing and filtering tool that is ideal for noisy financial time series. As discussed in \cite{huang2014hilbert}, the method of HHT and its variations have been applied in numerous fields, from engineering to geophysics.

Applications of HHT to finance date back to the work by Huang and co-authors on modeling mortgage rate data (\cite{Huang2003}). The empirical mode decomposition (EMD) has been used for financial time series forecasting (\cite{emd_forecast,wang2017forecasting}) and for examining the correlation between financial time series (\cite{nava2018dynamic}). 

In terms of methodology, there have been several  studies on the variations and alternatives to EMD, including optimization-based methods (\cite{hou2011adaptive,hou2013data,hou2009variant,huang2013optimization}), ensemble empirical-mode decomposition (EEMD) for tackling mode mixing (\cite{wu2009ensemble}), and other noise-assisted approaches (\cite{yeh2010complementary}). For financial time series with high level of intrinsic noise, we apply the complementary ensemble empirical mode decomposition (CEEMD). Like EMD, CEEMD   decomposes any time series -- stationary or not --  into a number of intrinsic mode functions representing the local characteristics of the time series at different timescales, but the timescale separation is improved by resolving mode mixing in EMD (\cite{huang1999new}). The noise-assisted approach is also more robust to intrinsic noise in the data. CEEMD have been found to be useful for forecasting (\cite{niu2016novel,tang2015novel}) and signal processing (\cite{li2015gpr}).

In this paper, we present a novel method of applying adaptive complementary ensemble empirical mode decomposition (ACE-EMD) and Hilbert spectral analysis to study cryptocurrency price dynamics. As an emerging asset class, cryptocurrencies have a number of salient features compared to traditional equities, including significantly higher volatility and rapidly changing directional trends (\cite{kristoufek2015main,bariviera2017some,katsiampa2017volatility,phillips2018cryptocurrency,bouri2019modelling,hu2019cryptocurrencies,kang2019co,tiwari2019time,garcia2020transfer}).  Our  noise-assisted approach decomposes any time series into a number of intrinsic mode functions, along with the corresponding instantaneous amplitudes and instantaneous frequencies. The decomposition is adaptive to the time-varying volatility of each cryptocurrency price evolution. Different combinations of modes allow us to reconstruct the time series using components of different timescales. We then apply Hilbert spectral analysis to define and compute the associated instantaneous energy-frequency spectrum to illustrate the properties of various timescales embedded in the original time series.  In particular, we derive and compute the central frequency associated with a collection of cryptocurrencies as well as equity indices, which allows us to observe the distinct behaviors of cryptocurrency prices.

The rest of the paper is structured as follows. First, we  present and implement the method of ACE-EMD in Sect. \ref{sect-hht}. Further applications to filtering cryptocurrency prices and estimating conditional volatilities are discussed in Sect. \ref{sect-scale}.  In Sect. \ref{sect-sepctrum}, we derive  the energy-frequency spectra of various cryptocurrencies and compare them to traditional equity indices. Lastly, concluding remarks are provided in Sect. \ref{sect-conclude}.


\section{Multiscale Decomposition Methodology}\label{sect-hht}

The method of empirical mode decomposition (EMD)  decomposes a time series $x(t)$ iteratively into a finite sequence of oscillating components $c_1(t), \cdots, c_n(t)$, plus a nonoscillatory trend called the residual term $r_n(t)$. Precisely, we have
\begin{equation}\label{eq-emd}
    x(t) = \sum_{j=1}^n c_j(t) + r_n(t).
\end{equation}

The components $c_j(t)$'s are called intrinsic mode functions (IMFs) as introduced by \cite{Huang1998}. They have certain oscillatory properties and admit well-behaved and physically meaningful Hilbert transform.  For each IMF, the numbers of extrema and  zero crossings must be equal or at most differ by one, and  the maxima of the function defined by the upper envelope and the minima defined by the lower envelope must sum up to zero at any time. These conditions  ensure  pure oscillation while allowing time-varying frequency and amplitude. Mathematically, an IMF $c(t)$ admits the expression 
\begin{equation}\label{eq-af}
    c(t) = a(t)\cos(\theta(t)),
\end{equation} 
where $a(t) \geq 0$ is the instantaneous amplitude, and $\theta(t)$ is the phase function with $\theta'(t) \geq 0$.

Following the standard EMD algorithm (\cite{Huang1998,rilling2003empirical}), we apply a sifting process that decomposes any time series into a finite set of IMFs that oscillate on different timescales, plus a nonoscillatory residual term. The key idea of the method is as follows: look for the finest oscillation by finding all the local maxima and minima, and then subtract the remaining trend, until the oscillation satisfies the IMF conditions. Each IMF discovered is removed sequentially from the time series until  a nonoscillatory residual term remains. The residual term is a constant or monotonic function, or has at most one maximum or minimum. 


  

When  decomposing a time series into   IMF components of different frequencies, the phenomenon of mode mixing may arise.  Mode mixing is defined as either one IMF consisting of widely disparate scales, or signals of similar scales residing in several IMF components  (\cite{huang1999new,wu2009ensemble}). This problem poses potential challenges on the interpretation of IMFs, and can be exacerbated by the  high degrees of nonstationarity and noise  commonly observed in cryptocurrency prices.  

 The ensemble empirical mode decomposition (EEMD)  was proposed by \cite{wu2009ensemble} to  resolve the mode mixing issue. It is a noise-assisted signal processing technique that extracts each component from an ensemble mean computed based on $N$ trials. In each trial $i$, an i.i.d. white noise $w_i(t)$ with a zero mean and finite variance is added to the original time series $x(t)$, and $x(t) + w_i(t)$ is referred to as the \emph{signal} in this trial. The plain EMD algorithm is then applied to the signal, outputting the IMFs $c_{ij}(t),\ j = 1,\cdots,n$, and the residual term $r_{in}(t)$. Finally, the ensemble mean of the IMF components and residual terms across all the $N$ trails is regarded as the true mode extraction. The components are given by
\begin{equation}
    c_j(t) = \frac{1}{N}\sum_{i=1}^N c_{ij}(t),  \quad r_n = \frac{1}{N}\sum_{i=1}^N r_{in}(t).
\end{equation}
Applying these to  \eqref{eq-emd}, we obtain 
\begin{equation}
    \sum_{j=1}^n c_{ij}(t) + r_{in}(t) = x(t) + w_i(t). \label{crxw}
\end{equation}
Summing up \eqref{crxw} over the trials divided by $N$, the ensemble mean of the IMF components is $x(t) + \frac{1}{N}\sum_{i=1}^N w_i(t)$, which converges to the original time series $x(t)$ almost surely at the rate of $\mathcal{O}(\frac{1}{\sqrt{N}})$. 



Hence, a large ensemble size is desired though it can also be prohibitive in terms of computational cost and speed. To address this particular issue, \cite{yeh2010complementary} introduce the complementary ensemble EMD (CEEMD) method. CEEMD adds a complementary negative noise $-w_i(t)$ to the ensemble for each trial, thus expanding the total ensemble size to $2N$.  The components from the ensemble mean  sum up to equal the original time series:
\begin{equation}\label{eq-ceemd}
     \sum_{j=1}^n c_j(t) + r_n(t) = x(t) + \frac{1}{2N}\sum_{i=1}^N (w_i(t) - w_i(t))= x(t).
\end{equation}
This holds exactly regardless of the choice of $N$, thus reducing the need to have a very large ensemble size. In addition to  reducing mode mixing,   CEEMD is also more robust to intrinsic noise in the original data, as it automatically averages out extra independent noises in the process. 

Both EEMD and CEEMD assume white noise with a  {constant variance} over time. Most financial time series, however, have nonstationary or time-varying volatility (see Fig. \ref{fig_vol}  below). This phenomenon is commonly observed among cryptocurrencies. Many cryptocurrencies tend to have a very low volume immediately after inception or initial coin offering (ICO), and some may experience huge rise in trading volume and volatility later. A constant noise level will decrease the signal to noise ratio when the time series isn't volatile, and produce a meaningless signal there. 

In order to capture the time-varying volatility of cryptocurrencies, we consider the \textit{adaptive complementary ensemble empirical mode decomposition} (ACE-EMD), where the noise level of $w_i(t)$ is proportional to the intrinsic noise level. Inspired by \cite{Huang2003}, we implement a pilot sifting process on $x(t)$ and take the first IMF component $c_p(t)$ for noise estimation. The full algorithm is summarized as follows:
\begin{itemize}
\item Implement a pilot sifting process on $x(t)$ and extract the first IMF component $c_p(t)$.
\item Estimate the local maxima of $c_p(t)$ and interpolate it with cubic spline $a_p(t)$. According to the definition of IMF, $c_p(t)$ is symmetric, so the upper envelope $a_p(t)$ estimates the amplitude of the pilot mode $c_p(t)$. 
\item For $i=1,\cdots,N$:
\begin{itemize}
\item Generate noise $w_i(t)$ from a distribution (e.g. Gaussian noise) such that \begin{equation}
\mathbb{E}[w_i(t)] = 0,\ Var[w_i(t)] = \sigma^2 a_1^2(t), \ w_i(s) \perp\mkern-10mu\perp w_i(t) \ \mbox{for} \ s\neq t.
\end{equation}
\item Implement EMD on $x(t) + w_i(t)$ and decompose it into IMF components $c_{i1}^+(t), \cdots, c_{in}^+(t)$ plus a residual term $r_{in}^+(t)$. 
\item Implement EMD on $x(t) - w_i(t)$ and decompose it into IMF components $c_{i1}^-(t), \cdots, c_{in}^-(t)$ plus a residual term $r_{in}^-(t)$. 
\end{itemize}
\item Compute the decomposition of $x(t)$ as the ensemble mean: \begin{align*}
 c_j(t) &= \frac{1}{2N}\sum_{i=1}^N \left( c_{ij}^+(t) + c_{ij}^-(t)\right), \quad j = 1,\cdots,n, \\
 r_n(t) &= \frac{1}{2N}\sum_{i=1}^N \left( r_{in}^+(t) + r_{in}^-(t)\right).
\end{align*}
\end{itemize}
Applying these to  \eqref{eq-ceemd}, we conclude that the components from the ensemble mean sum up to the original time series, so ACE-EMD ensures an exact decomposition. 

It remains to determine the only parameter in the algorithm, the noise level $\sigma$. To that end we choose $\sigma$ to optimize the orthogonality and separability of the decomposition (see e.g. \cite{yang2018causal}). Following \cite{Huang1998}, we define the \emph{orthogonality index} of the decomposition as 
\begin{equation}
OI = \frac{1}{T}\int_0^T \sum_{j \neq k} \frac{c_j(t) c_k(t)}{x^2(t)} dt,
\end{equation}
where the we sum over all the mode pairs, including the residual term $r_n(t)$ as the $(n+1)$-th mode. Similarly, the separability is defined as the root-mean-square of the pairwise correlation between the modes (see \cite{yang2018causal}). The strategy is to choose $\sigma$ minimizing the separability while keeping the absolute value of the orthogonality index small (e.g. $|OI |<0.05$).

\begin{figure}[t]
  \centering  
    \includegraphics[width=3.2in]{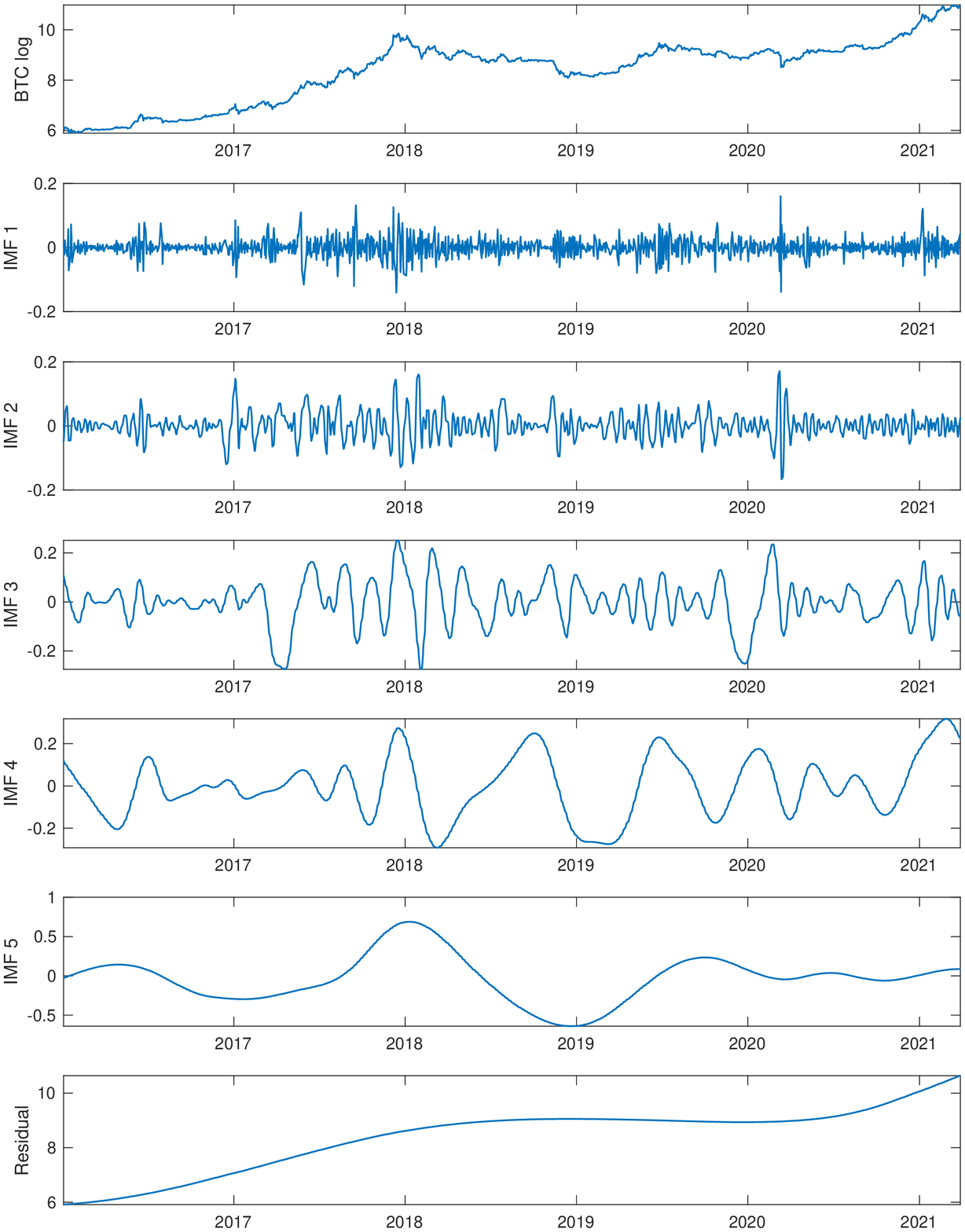}
    \includegraphics[width=3.2in]{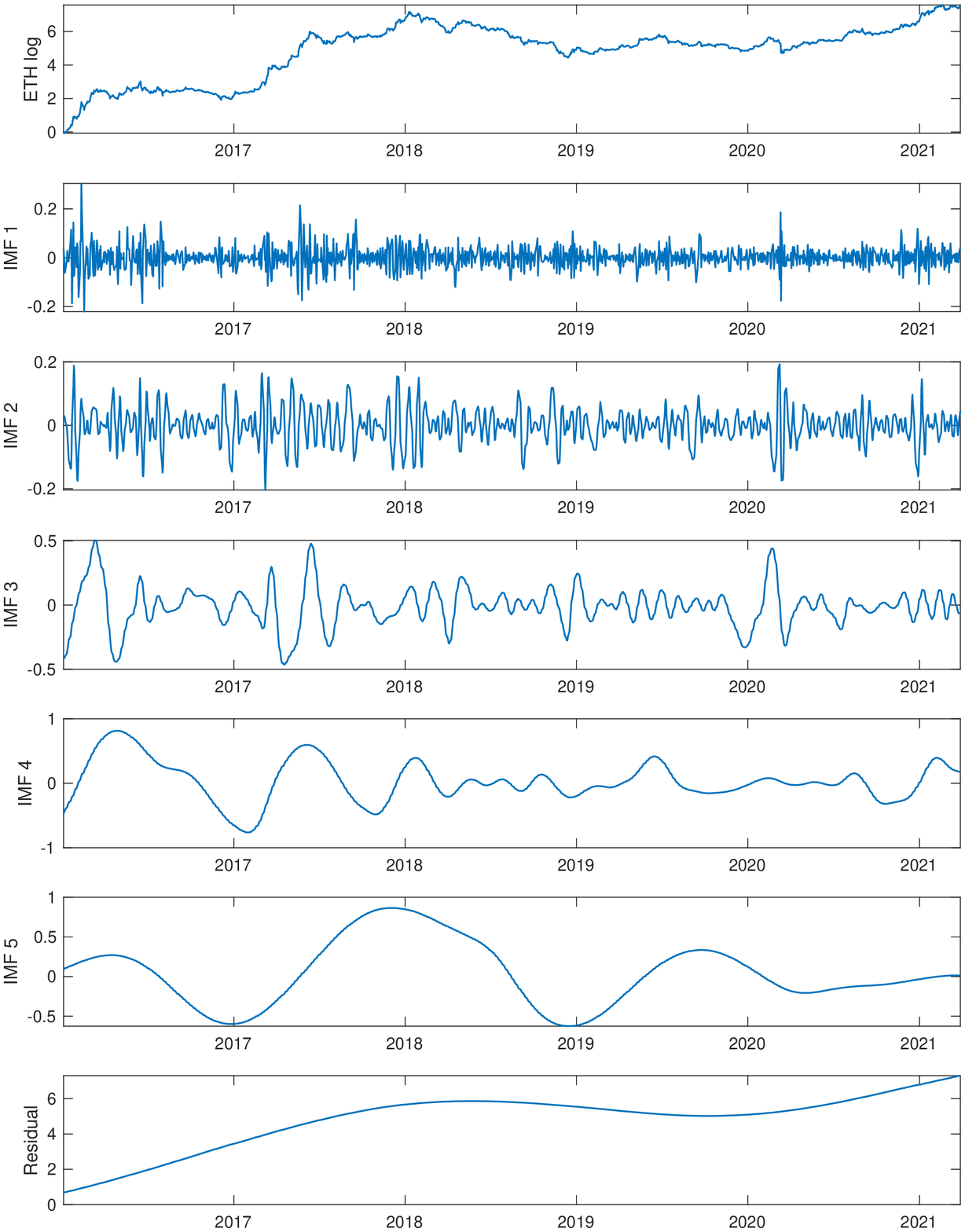}
    \caption{\small{IMFs and residual terms extracted from ACE-EMD of BTC (left) and ETH (right), from 1/4/2016 to 3/29/2021. The top row in each plot shows the original time series (in log prices). The subsequent rows show the IMFs of the corresponding time series, except for the bottom row which shows the residuals.}}\label{fig_eemd}
\end{figure}

In Fig. \ref{fig_eemd}, we show the decomposition from ACE-EMD for the log prices of  BTC and ETH . The top row in each plot shows the original time series $x(t)$, followed by the IMF $c_j(t)$'s with decreasing frequencies, ended with the residual term.  The number of IMF components $n=5$, so along with the residual term, there are 6 components in total.  As we can see, the first IMF exhibits the highest frequency of fluctuation, whereas the smooth residual term reflects the overall trend.

\section{Timescale Separation with ACE-EMD}\label{sect-scale}
Financial time series often exhibit characteristics on different timescales. The short-term scale explains more of the volatility while the long-term scale reflects the trend. Separating a time series onto different timescales provides extra insights of the behavior and dynamics of the process, and ACE-EMD naturally serves as a filter bank and allows reconstruction onto different scales.
 
 \subsection{Time Series Filtering and Reconstruction}\label{scet_recon}
By construction of the EMD algorithm, the iterative sifting process identifies the finest structures, and then extract longer and longer scales. As a result, the first few components have higher frequency which are more noisy, and the last few components have lower frequency representing long term structure. Hence, ACE-EMD can be used as a filter for time series. 

In the reconstruction of the original time series using the IMF components, we can choose a subset of modes as a filter for desired information. By removing the first few high frequency components, we create a low-pass filter; that is,
\begin{equation}
    x_L^{(m_l)}(t) =  x(t) - \sum_{j=1}^{n-m_l+1} c_j(t). \label{lowpass}
\end{equation}
This reconstruction using only the last few components can serve as a smoothing of the time series. Similarly, we can also build a high-pass filter with
\begin{equation}
    x_H^{(m_h)}(t) = \sum_{j=1}^{m_h} c_j(t), \label{highpass}
\end{equation}
which captures the high-frequency local behaviors, and can also be used to estimate the noise level or volatility. 

In each case, the number of components (including the residual term) equals to $m_l$ and $m_h$ respectively. We use $x_L^{(m_l)}(t)$ and $x_H^{(m_h)}(t)$ to denote the low-pass and high-pass filtered reconstruction of $x(t)$ with $m_l$ and $m_h$ components. Note that the low-pass filter and the high-pass filter are complementary to each other when $m_l + m_h = n+1$, where $n+1$ is the total number of components (including the residual term).

\begin{figure}[ht]
   \centering
      \includegraphics [width=6.8in]
{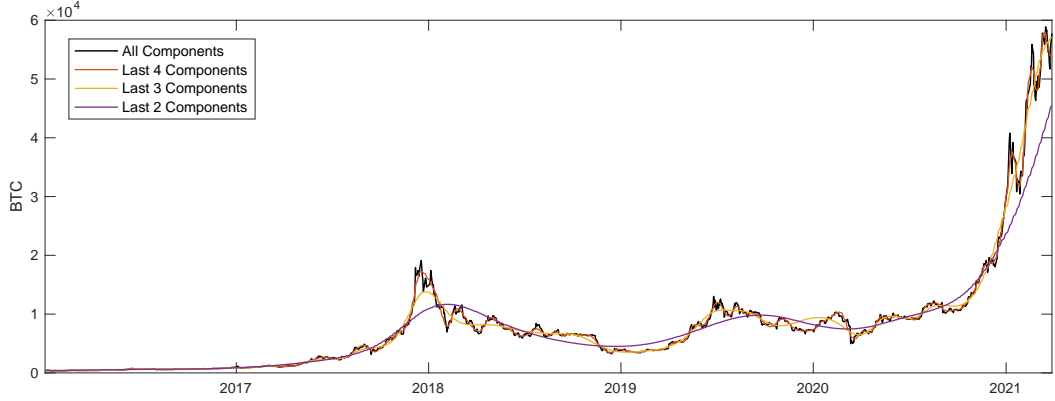} 
\caption{\small{Reconstruction of BTC price (in \$) time series  from 1/4/2016 to 3/29/2021. The decomposition is implemented on the log price using the low-pass filter in \eqref{lowpass}, and then exponentiated to show the reconstruction in US dollar.}}
    \label{fig_recon}
\end{figure}

In Fig. \ref{fig_recon}, we illustrate the low-pass filtered reconstruction of BTC and ETH prices. This involves applying \eqref{lowpass} using different collections of components. Specifically, we have used the last 4, 3, and 2 components including the residual term, i.e. $x_L^{(m_l)}(t)$ with $m_l = 4,3,2$. The reconstructed log prices are taken exponential to approximate the original price data. We can see that, with more components included  in the reconstruction, the resulting time series resembles the original time series on a finer  timescale.  Compared to some other time series smoothing techniques, such as moving average, the ACE-EMD low-pass filter achieves smoothing without lags. 

While the low-pass filter with ACE-EMD exhibits smoothed trends in the time series, the high-pass filter removes nonstationarity thus making the time series ready for statistical analysis. Table \ref{table:stationary} shows the stationarity test results for BTC, ETH, XRP and LTC, before and after high/low-pass filtering. We experimented with three stationarity (unit-root) tests, namely the Augmented Dickey-Fuller (ADF) test (\cite{cheung1995lag}), the Kwiatkowski, Phillips, Schmidt, and Shin (KPSS) test (\cite{kwiatkowski1992testing}), and the Phillips-Perron (PP) test (\cite{phillips1988testing}). The results show that the highly nonstationary cryptocurrency prices become stationary after high-pass filtered with ACE-EMD. The three stationarity tests suggest consistently that a high-pass filter with one or two ACE-EMD components is favorable when doing timescale separation.

\begin{table}
\centering
\begin{scriptsize}
\begin{tabular}{c|c c c c c|c|c c c c c}
\hline
\textbf{BTC} & $x_H^{(1)}$ & $x_H^{(2)}$ & $x_H^{(3)}$ & $x_H^{(4)}$ & $x_H^{(5)}$ & $x$ & $x_L^{(5)}$ & $x_L^{(4)}$ & $x_L^{(3)}$ & $x_L^{(2)}$ & $x_L^{(1)}$ \\
\hline
ADF test & \textbf{0.0010} & \textbf{0.0010} & \textbf{0.0010} & \textbf{0.0010} & \textbf{0.0172} & 0.9985 & 0.9990 & 0.9990 & 0.9990 & 0.9990 & 0.9990 \\ 
KPSS test & \textbf{0.1000} & \textbf{0.1000} & 0.0100 & 0.0100 & 0.0100 & 0.0100 & 0.0100 & 0.0100 & 0.0100 & 0.0100 & 0.0100 \\
PP test & \textbf{0.0010} & \textbf{0.0010} & \textbf{0.0010} & \textbf{0.0010} & \textbf{0.0172} & 0.9985 & 0.9990 & 0.9990 & 0.9990 & 0.9990 & 0.9990 \\
\hline
\hline
\textbf{ETH} & $x_H^{(1)}$ & $x_H^{(2)}$ & $x_H^{(3)}$ & $x_H^{(4)}$ & $x_H^{(5)}$ & $x$ & $x_L^{(5)}$ & $x_L^{(4)}$ & $x_L^{(3)}$ & $x_L^{(2)}$ & $x_L^{(1)}$ \\
\hline             
ADF test & \textbf{0.0010} & \textbf{0.0010} & \textbf{0.0010} & \textbf{0.0010} & \textbf{0.0168} & 0.9840 & 0.9990 & 0.9990 & 0.9990 & 0.9990 & 0.9990 \\                                                                 
KPSS test & \textbf{0.1000} & \textbf{0.1000} & 0.0104 & 0.0100 & 0.0100 & 0.0100 & 0.0100 & 0.0100 & 0.0100 & 0.0100 & 0.0100 \\                                                                                       
PP test & \textbf{0.0010} & \textbf{0.0010} & \textbf{0.0010} & \textbf{0.0010} & \textbf{0.0168} & 0.9840 & 0.9990 & 0.9990 & 0.9990 & 0.9990 & 0.9990 \\     
\hline 
\hline           
\textbf{XRP} & $x_H^{(1)}$ & $x_H^{(2)}$ & $x_H^{(3)}$ & $x_H^{(4)}$ & $x_H^{(5)}$ & $x$ & $x_L^{(5)}$ & $x_L^{(4)}$ & $x_L^{(3)}$ & $x_L^{(2)}$ & $x_L^{(1)}$ \\
\hline       
ADF test & \textbf{0.0010} & \textbf{0.0010} & \textbf{0.0010} & \textbf{0.0010} & \textbf{0.0131} & \textbf{0.0263} & \textbf{0.0012} & \textbf{0.0010} & \textbf{0.0010} & \textbf{0.0010} & \textbf{0.0010} \\      
KPSS test & \textbf{0.1000} & \textbf{0.1000} & \textbf{0.1000} & 0.0100 & 0.0100 & 0.0100 & 0.0100 & 0.0100 & 0.0100 & 0.0100 & 0.0100 \\                                                                                          
PP test & \textbf{0.0010} & \textbf{0.0010} & \textbf{0.0010} & \textbf{0.0010} & \textbf{0.0131} & \textbf{0.0263} & \textbf{0.0012} & \textbf{0.0010} & \textbf{0.0010} & \textbf{0.0010} & \textbf{0.0010} \\ 
\hline 
\hline       
\textbf{LTC} & $x_H^{(1)}$ & $x_H^{(2)}$ & $x_H^{(3)}$ & $x_H^{(4)}$ & $x_H^{(5)}$ & $x$ & $x_L^{(5)}$ & $x_L^{(4)}$ & $x_L^{(3)}$ & $x_L^{(2)}$ & $x_L^{(1)}$ \\
\hline       
ADF test & \textbf{0.0010} & \textbf{0.0010} & \textbf{0.0010} & \textbf{0.0010} & \textbf{0.0235} & 0.9325 & 0.9971 & 0.9990 & 0.9990 & 0.9990 & 0.9990 \\     
KPSS test & \textbf{0.1000} & \textbf{0.1000} & \textbf{0.1000} & 0.0100 & 0.0100 & 0.0100 & 0.0100 & 0.0100 & 0.0100 & 0.0100 & 0.0100 \\                      
PP test & \textbf{0.0010} & \textbf{0.0010} & \textbf{0.0010} & \textbf{0.0010} & \textbf{0.0235} & 0.9325 & 0.9971 & 0.9990 & 0.9990 & 0.9990 & 0.9990 \\                                                                  
\hline        
\end{tabular}
\caption{The  $p$-values from  stationarity tests of the low/high-pass filtered time series of four major cryptocurrency prices. Results from  the high-pass filtered time series are on the left with increasing number of components included; results from the original unfiltered time series $x$ are placed in the middle column, and results from the low-pass filtered time series are on the right with decreasing number of components included. The null hypotheses of ADF test and PP test are nonstationary (unit-root), while the null hypothesis of KPSS test is stationary (no unit-root). The bold-highlighted values indicate stationarity under 0.05 significant level.}
\label{table:stationary}
\end{scriptsize}
\end{table}

\clearpage

\subsection{Multiscale Volatilities and Correlations}

To analyze the long-term and short-term behaviors of the time series, we apply \eqref{lowpass} and \eqref{highpass} to filter the original time series into low-pass and high-pass components. Statistical analysis such as mean and volatility can be defined on the filtered time series. To set up, let $s(t)$ be the price of the asset price, and $x(t) = \log s(t)$ be the log price. Implement ACE-EMD on $x(t)$ and apply \eqref{lowpass} and \eqref{highpass} to get $x_L^{(m_l)}(t)$ and $x_H^{(m_h)}(t)$, with $m_l$ and $m_h$ being the number of components in the low-pass and high-pass filtering respectively. To study the statistical properties of the asset prices, we define the low-pass and high-pass log returns:
\begin{align}\label{eq-return}
r_L^{(m_l)}(t) &= x_L^{(m_l)}(t) - x_L^{(m_l)}(t-1), \\ r_H^{(m_h)}(t) &= x_H^{(m_h)}(t) - x_H^{(m_h)}(t-1).
\end{align}
The low-pass and high-pass volatility are   the standard deviation of the low-pass and high-pass returns, defined by the unbiased estimators
\begin{align}
\sigma_L^{(m_l)} &= \sqrt{\frac{1}{T-1}\sum_{t=1}^T \left( r_L^{(m_l)}(t) - \mu_L^{(m_l)}\right)^2},\\ \quad \sigma_H^{(m_h)} &=  \sqrt{\frac{1}{T-1}\sum_{t=1}^T \left( r_H^{(m_h)}(t) - \mu_H^{(m_h)}\right)^2} \,,
\end{align}
where 
\begin{equation}
\mu_L^{(m_l)} = \frac{1}{T}\sum_{t=1}^T r_L^{(m_l)}(t), \qquad \mu_H^{(m_h)} = \frac{1}{T}\sum_{t=1}^T r_H^{(m_h)}(t)
\end{equation} are the mean low-pass and high-pass log returns. 

%

\begin{figure}[ht]
   \centering
    \includegraphics[width=6.2in]{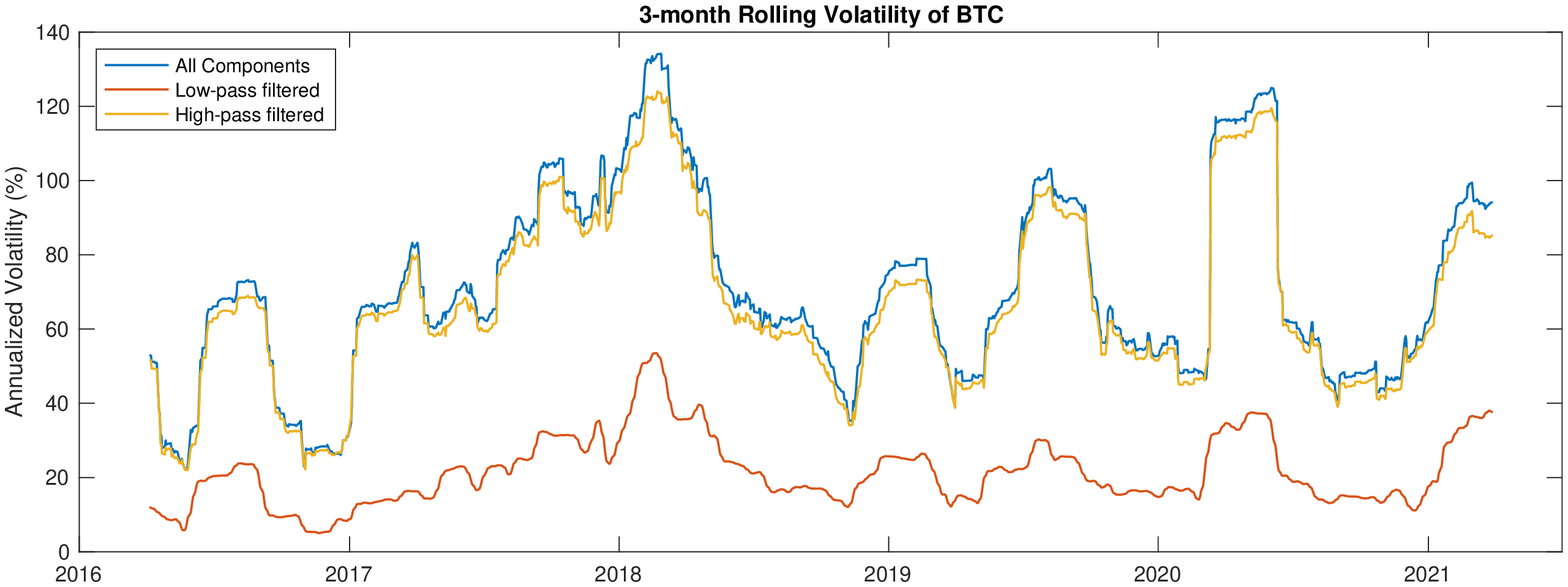} \\
        \includegraphics[width=6.2in]{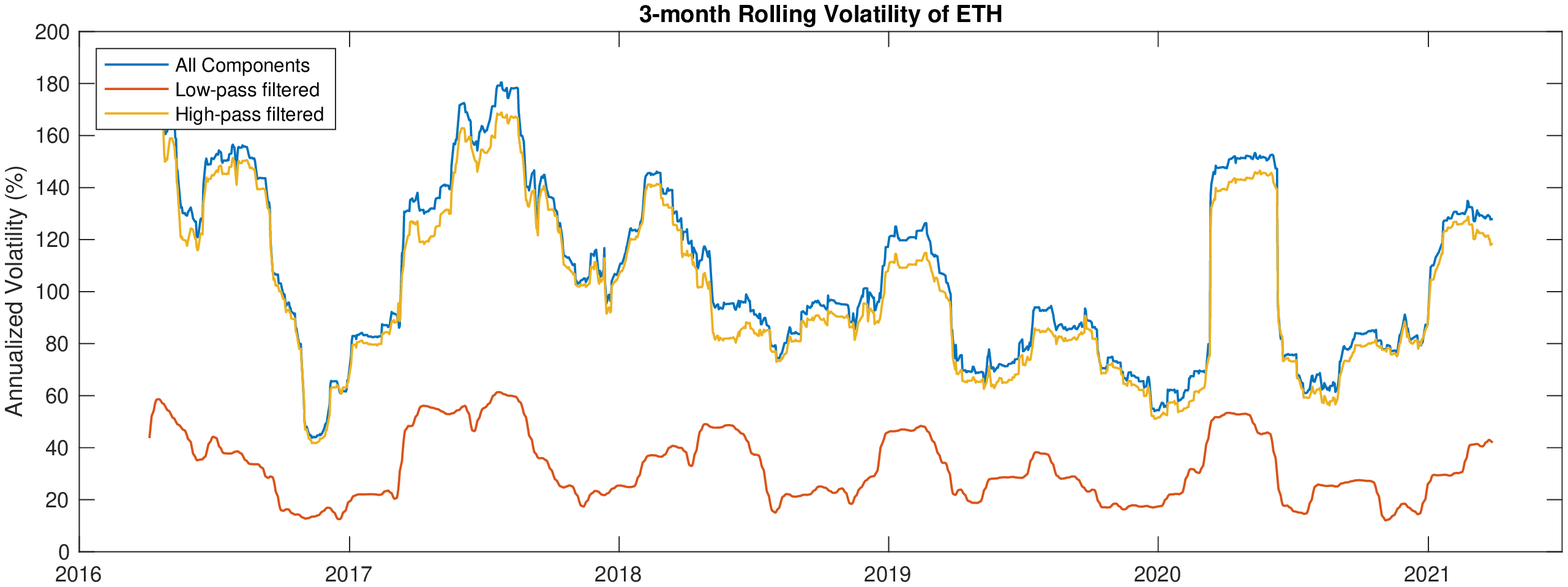} \\
        \includegraphics[width=6.2in]{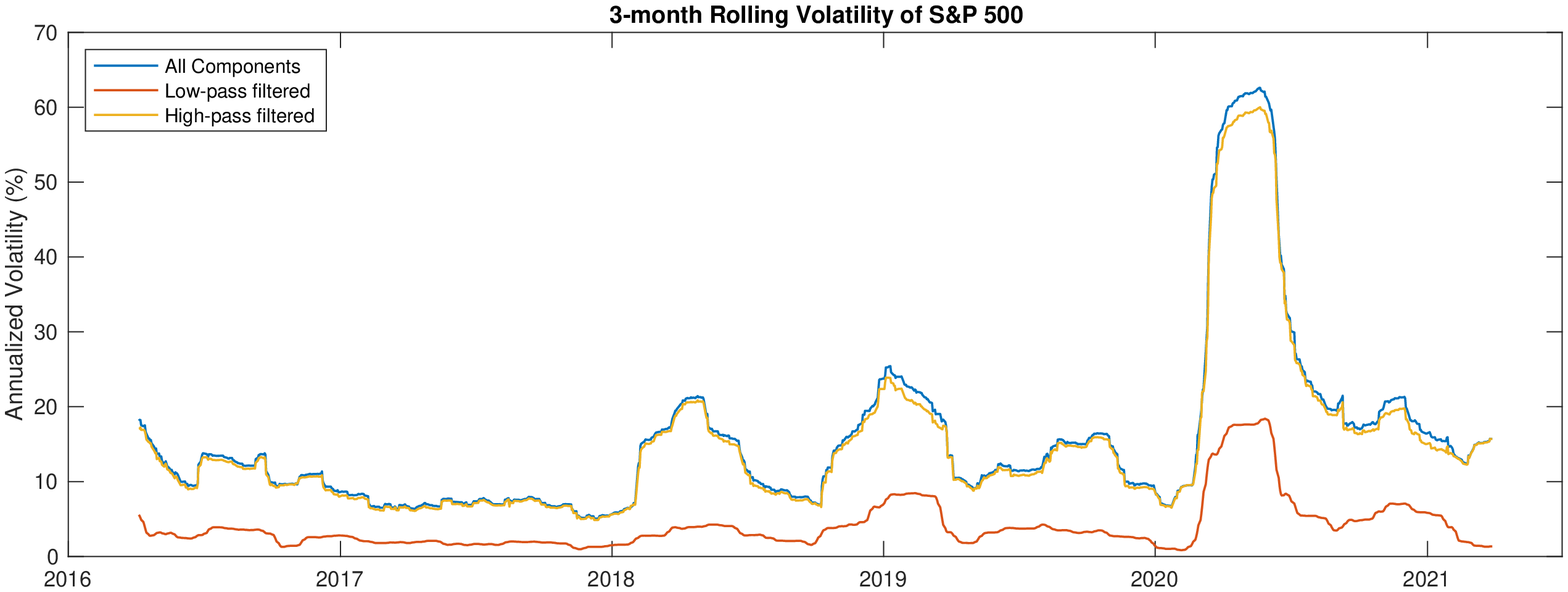} 
\caption{\small{3-month rolling volatility of BTC, ETH, and S\&P 500 from 1/4/2016 to 3/29/2021. Values are computed using all components, high-pass ($m_h=2$) filtered data, and low-pass ($m_l=4$) filtered data. }}
    \label{fig_vol}
\end{figure}

Fig. \ref{fig_vol} shows the 3-month rolling volatility of BTC, ETH, and S\&P 500 computed using all components, high-pass filtered data, and low-pass filtered data. Recall from  Fig. \ref{fig_eemd} that the time series are decomposed into 6 components including the residual terms. We use $m_l = 4$ for low-pass filtering and $m_h=2$ for high-pass filtering. Notice that the high-pass filtered time series captures most of the volatility residing in the original times series. Therefore, we consider using the stationary high-pass filtered data for statistical analysis.

 It has been observed in the financial market that the volatility of asset returns is usually asymmetric, i.e. the the volatility is higher in a downside market than that in an upside market. Following \cite{bekaert2000asymmetric}, we capture the asymmetry by looking at the conditional volatility. Specifically, we examine if the asymmetry exists in the price dynamics. To that end, we define the conditional volatilities based on the ACE-EMD high-pass filter as follows:
\begin{align}
\sigma_{+H}^{(m_h)} = \sqrt{Var(r_H^{(m_h)}(t) | r_H^{(m_h)}(t-1) > \mu_H^{(m_h)})}, \\
\sigma_{-H}^{(m_h)} = \sqrt{Var(r_H^{(m_h)}(t) | r_H^{(m_h)}(t-1) < \mu_H^{(m_h)})}.
\end{align}
In essence, $\sigma_{+H}^{(m_h)}$ and $\sigma_{-H}^{(m_h)}$ capture the high-pass volatilities conditioned on an upside movement and a downside movement respectively.

\begin{figure}[ht]
   \centering
    \includegraphics[width=6.2in]{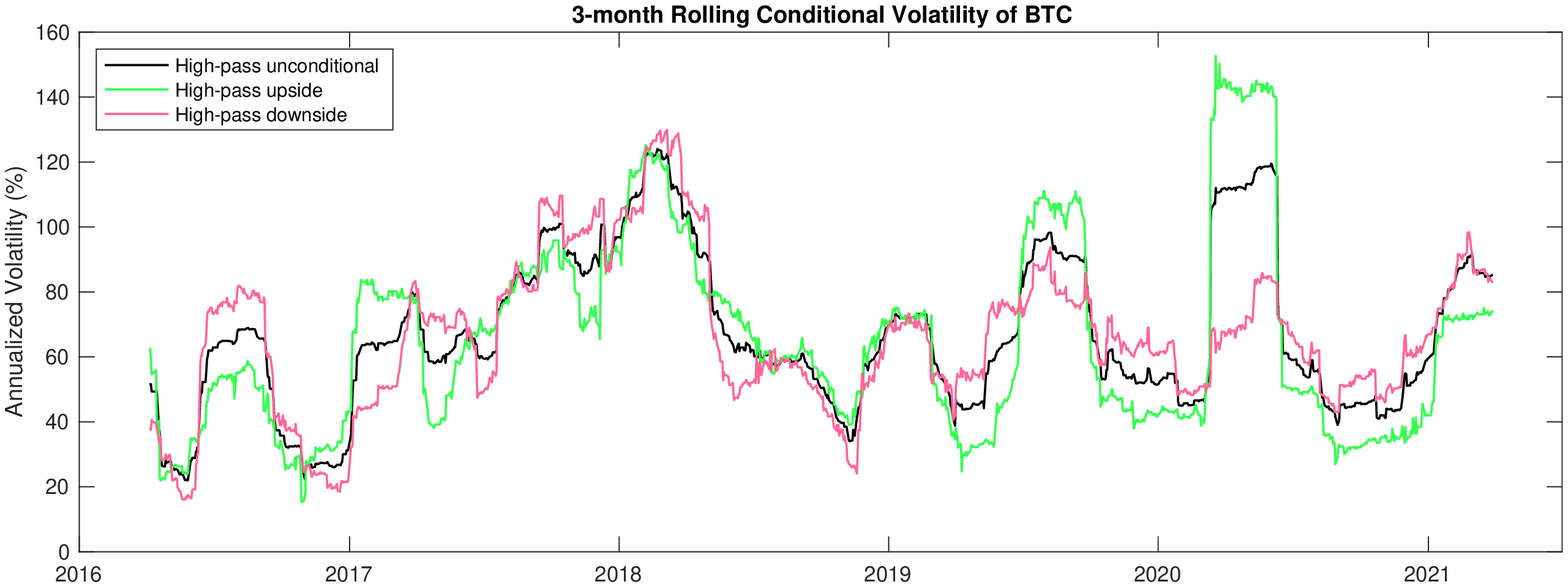} \\
        \includegraphics[width=6.2in]{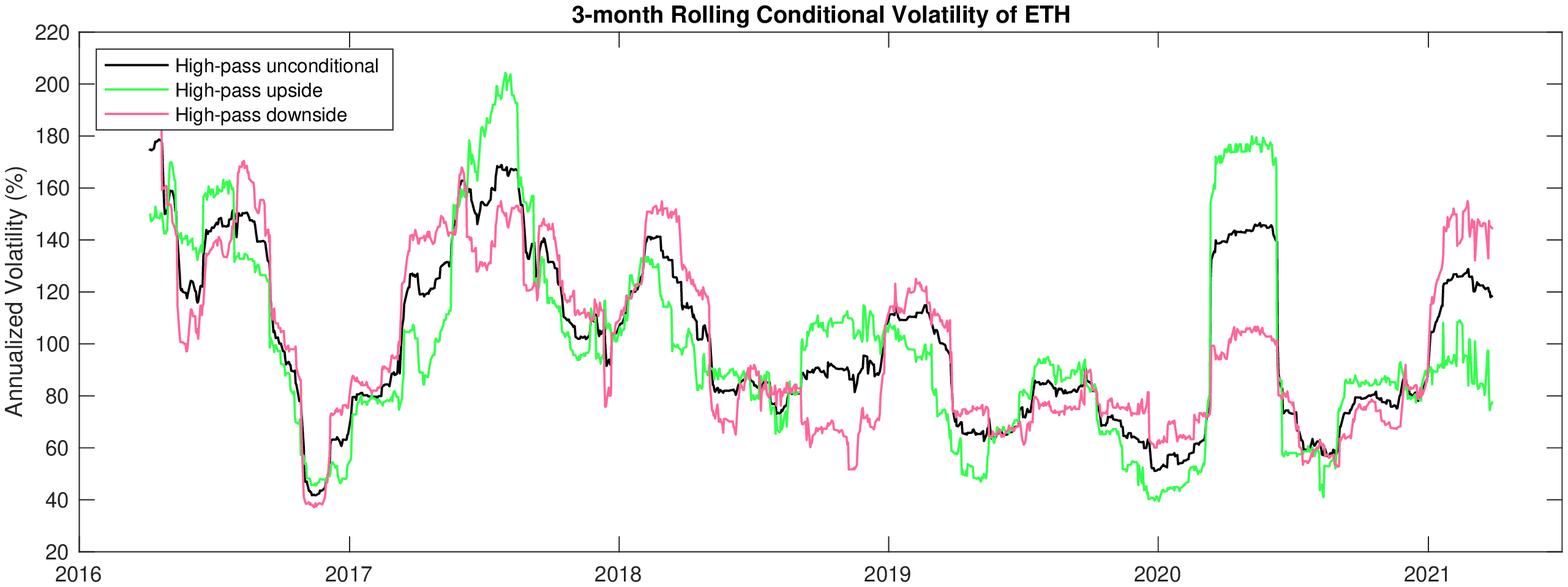} \\
        \includegraphics[width=6.2in]{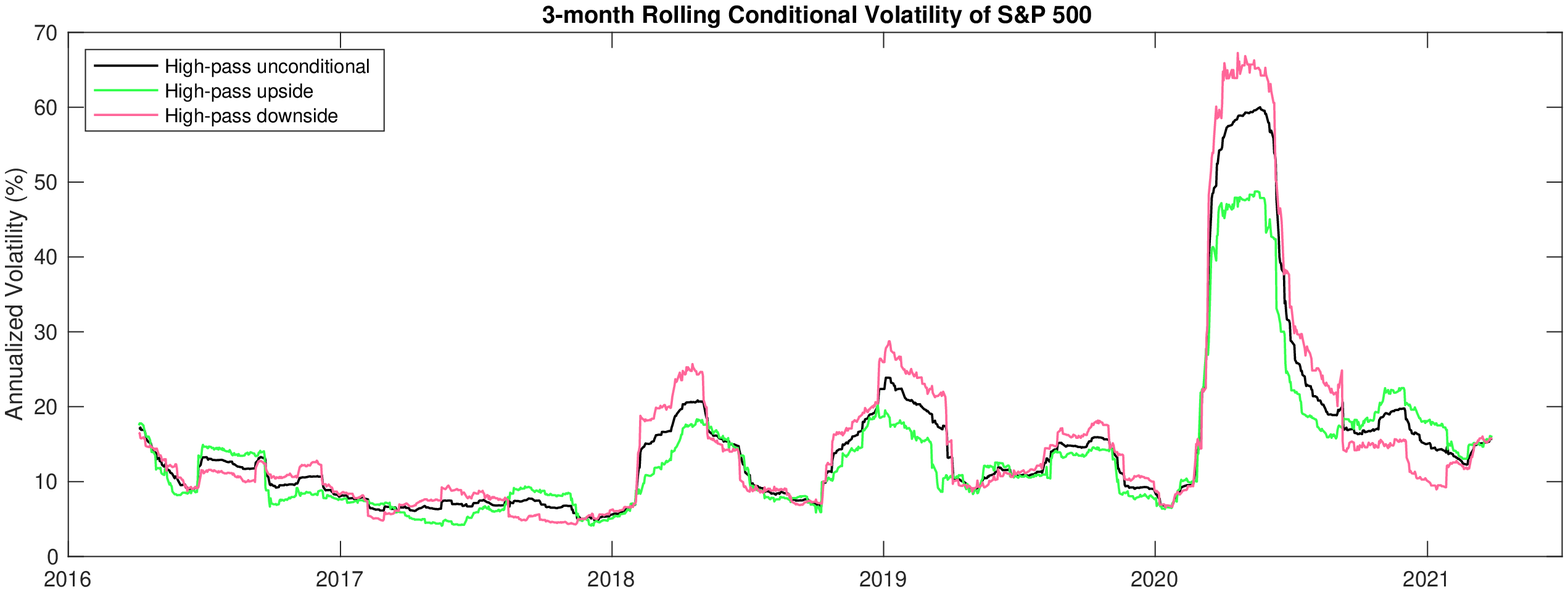} 
\caption{\small{3-month rolling estimation of the conditional volatility using low-pass and high-pass filtered data of BTC, ETH, and S\&P 500 from 1/4/2016 to 3/29/2021. Values are computed using high-pass ($m_h=2$) filtered data. }}
    \label{fig_convol}
\end{figure}

Fig. \ref{fig_convol} shows the 3-month rolling estimation of the conditional volatility using high-pass filtered data of BTC, ETH, and S\&P 500. In the high-pass components, we see that S\&P 500 shows asymmetric volatility with a larger downside volatility most of the time. The cryptocurrencies, however, exhibit both directions of asymmetry, where upside and downside movements trigger high volatility alternately during different periods of time. This phenomenon suggests a distinct behavior of the cryptocurrencies comparing with the traditional equity indices.

\clearpage

Define the events of upside volatility asymmetry and downside volatility asymmetry as:
\begin{equation}\label{asym_up}
A^+ = (\sigma_{+H}^{(m_h)} - \sigma_{-H}^{(m_h)}) > \epsilon \sigma_{H}^{(m_h)}
\end{equation}
\begin{equation}\label{asym_down}
A^- = (\sigma_{+H}^{(m_h)} - \sigma_{-H}^{(m_h)}) < -\epsilon \sigma_{H}^{(m_h)}
\end{equation}

\begin{figure}[ht]
   \centering
    \includegraphics[width=6in]{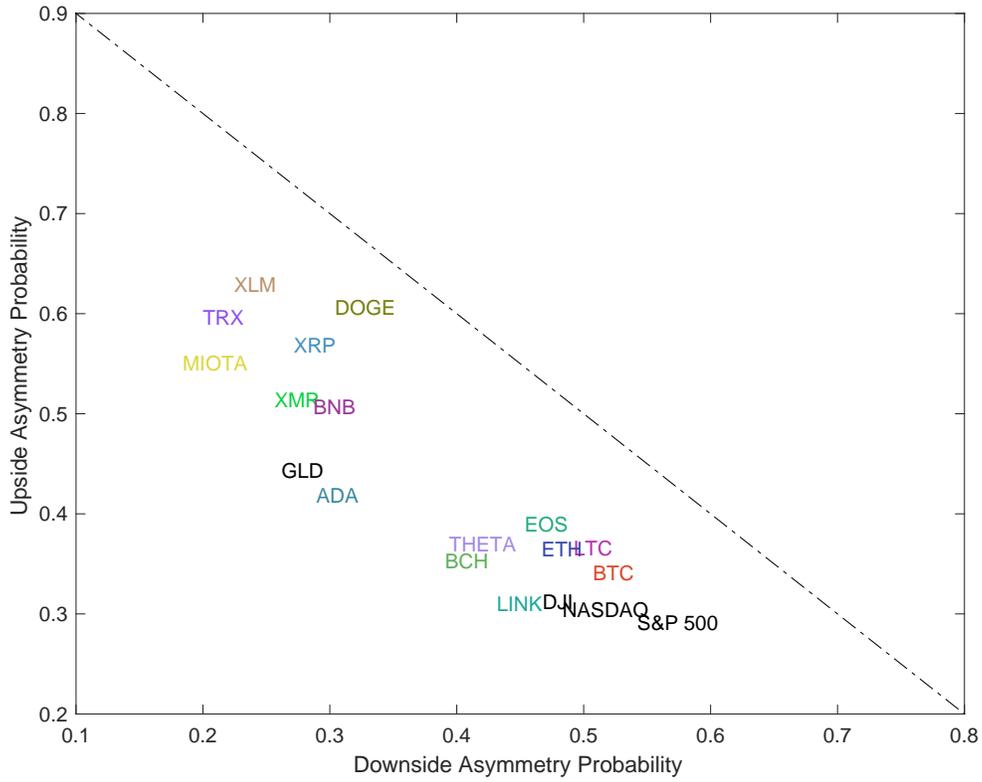} 
\caption{\small{Asymmetric volatility effect for cryptocurrencies and traditional financial securities, estimated using prices from 1/4/2016 to 3/29/2021. Points closer to the lower right corner are more dominated by downside volatility, while the upper left region shows more upside volatility. Points closer to the dashed line show more volatility asymmetry in general. }}
    \label{fig_asynvol}
\end{figure}

In Fig. \ref{fig_asynvol}, we plot the frequency of event $A^+$ as defined in \eqref{asym_up}, against the frequency of event $A^-$ as defined in \eqref{asym_down}. The dashed line corresponds to $\mathbb{P}(A^+) + \mathbb{P}(A^-) = 1$, The closeness to the dashed line indicates the level of volatility asymmetry in general. Towards the upper left direction the time series are upside volatility asymmetry dominated, while towards the lower right direction the time series are downside volatility asymmetry dominated. We can see there is a clear gap separating the assets into two clusters of upside and downside asymmetry. The three equity indices, S\&P 500, DJI, and NASDAQ, are all on the downside volatility asymmetry end. The top cryptocurrencies like BTC and ETH are also in that region. In contrast, the gold ETF (GLD) is in the upside volatility asymmetry region. Several cryptocurrencies, such as XRP, TRX, XLM, and DOGE,   are also found in the upside volatility asymmetry region.

\clearpage

\section{Energy-Frequency Spectrum}\label{sect-sepctrum}
Spectral analysis provides an alternative aspect to analyze time series on the frequency domain (\cite{koopmans1995spectral}). To that end, the definition of an IMF in HHT lends itself to  Hilbert spectral analysis. In this section, we apply this approach to compute and analyze the energy-frequency spectra of a wide collection of cryptocurrencies.

 \subsection{Hilbert Spectral Analysis}\label{sect-hsa}
 Recall that an oscillating real-valued function can be viewed as the projection of an orbit on the complex plane onto the real axis. For any function in time $X(t)$, the Hilbert transform is given by
\begin{equation}\label{hilbert}
    Y(t) = \mathcal{H}[X](t) := \frac{1}{\pi} \int_{-\infty}^{+\infty} \frac{X(s)}{t-s}\, ds,
\end{equation}
where the improper integral is defined as the Cauchy principal value. The transform exists for any function in $L^p$ (see \cite{titchmarsh1948introduction}). As a result, $Y(t)$ provides the complementary imaginary part of $X(t)$ to form an analytic function in the upper half-plane defined by
\begin{equation}
    Z(t) = X(t) + i Y(t) = a(t) e^{i\theta(t)},
\end{equation}
where
\begin{align}\label{ampEq}
    a(t) &= \|Z(t)\| = \sqrt{X^2(t) + Y^2(t)},\\
    \theta(t) &= \arg Z(t) = \arctan(\frac{Y(t)}{X(t)}). \label{phaseEq}
\end{align}

As shown by \cite{bedrosian,nuttall}, for a function of the form in \eqref{eq-af}, if the amplitude   and   frequency  are slow modulations, the Hilbert transform will give a $\pi/2$ shift to the phase $\theta(t)$. Therefore, the $a(t)$ given by \eqref{ampEq} is   the instantaneous amplitude, and the $\theta(t)$ given by \eqref{phaseEq} is   the instantaneous phase function. The instantaneous frequency is then defined as the $2\pi$-standardized rate of change of the phase function:
\begin{equation}\label{freqEq}
    f(t) = \frac{1}{2\pi}\dot{\theta}(t) = \frac{1}{2\pi}\frac{d}{dt}\left( \arctan(\frac{Y(t)}{X(t)}) \right).
\end{equation}


Applying Hilbert transform to each of the IMF components individually yields  a sequence of analytic signals (see \cite{Huang1998}):
\begin{equation}\label{eq_complex_imf}
    c_j(t) + i \mathcal{H}[c_j](t) =  a_j(t) e^{i\theta_j(t)},
\end{equation}
for $j = 1, \cdots, n$. In turn, the original time series can be represented as
\begin{equation}\label{eq-hs}
    x(t) = \mathfrak{Re}\ \sum_{j=1}^n a_j(t) e^{i \int^t 2\pi f_j(s) ds} + r_n(t).
\end{equation}
This decomposition can be seen as a sparse spectral representation of the time series with time-varying amplitude and frequency. In other words, each IMF represents a generalized Fourier expansion that are suitable for nonlinear and nonstationary financial time series. In summary, the procedure generates a series of complex functions that are analytic in time, along with their associated instantaneous amplitudes and instantaneous frequencies. These components capture different time scales and resolutions embedded in the time series and are used for time series filtering and reconstruction.

Lastly, the Hilbert spectrum is defined by
\begin{equation}
   H(f,t) = \sum_{j=1}^n H_j(f,t),  \quad \mbox{where}\ {\displaystyle H_{j}(f ,t)={\begin{cases}a_{j}(t),&f =f_{j}(t),\\0,&{\text{otherwise.}}\end{cases}}}
\end{equation}
The instantaneous energy of the $j$-th component is defined as
\begin{equation}\label{enrgEq}
E_j(t) = |a_j(t)|^2.
\end{equation}
We examine the behavior of the Hilbert spectrum through the pair $(f_j(t),E_{j}(t))$ for $t   \in [0,T]$ and $j = 1,\cdots,n$ (see Fig. \ref{fig:mode_spec}), which forms a sparse energy-frequency spectrum. Through this new lens,  we examine the behaviors of the time series.

\subsection{Central Frequency and Power Spectrum}

\begin{figure} [ht]
 \centering
    \includegraphics[trim = 0.5cm   0.5cm   1.3cm  0.5cm, clip,width=3.3in]{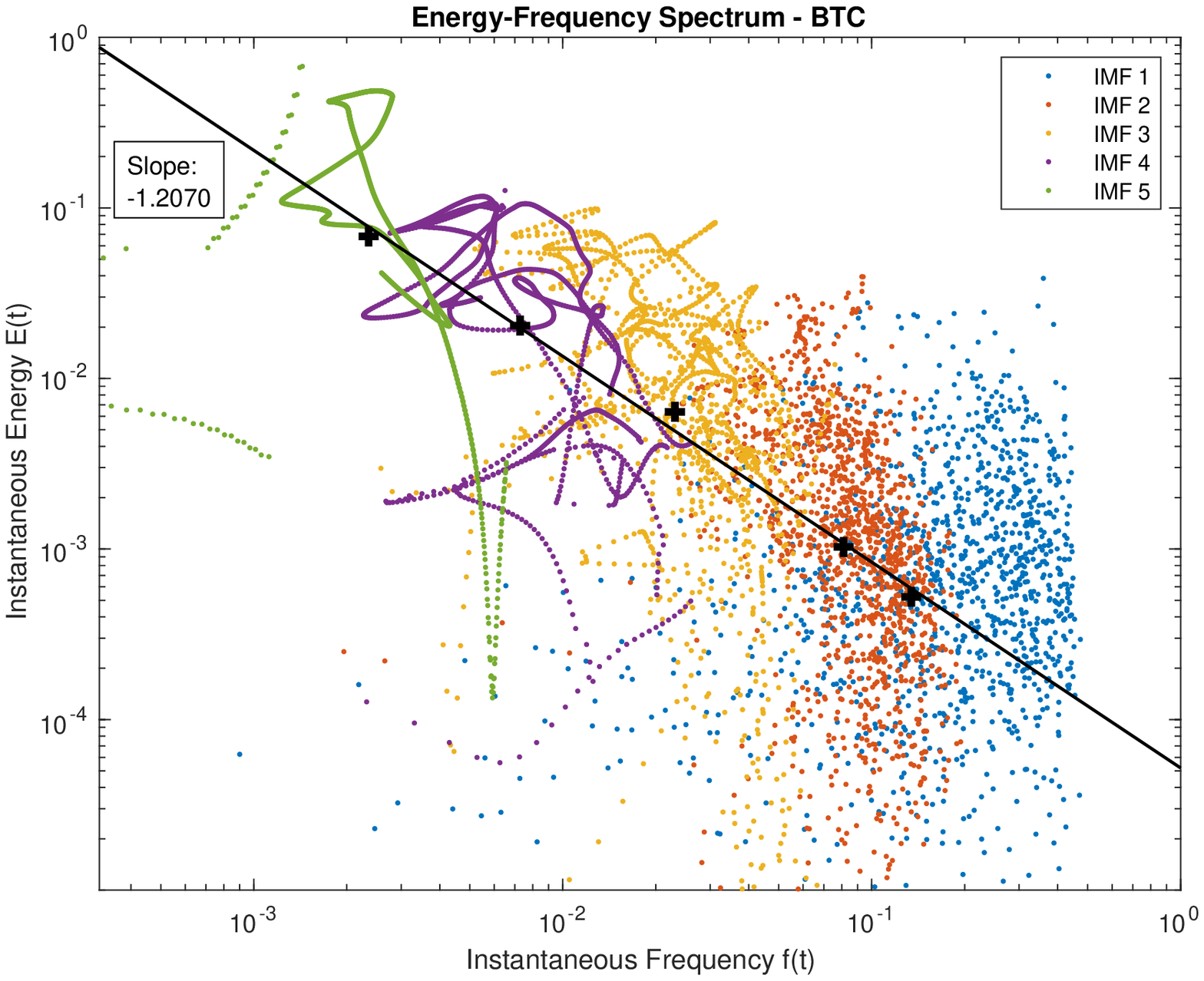}
        \includegraphics[trim = 0.5cm   0.5cm   1.3cm  0.5cm, clip,width=3.3in]{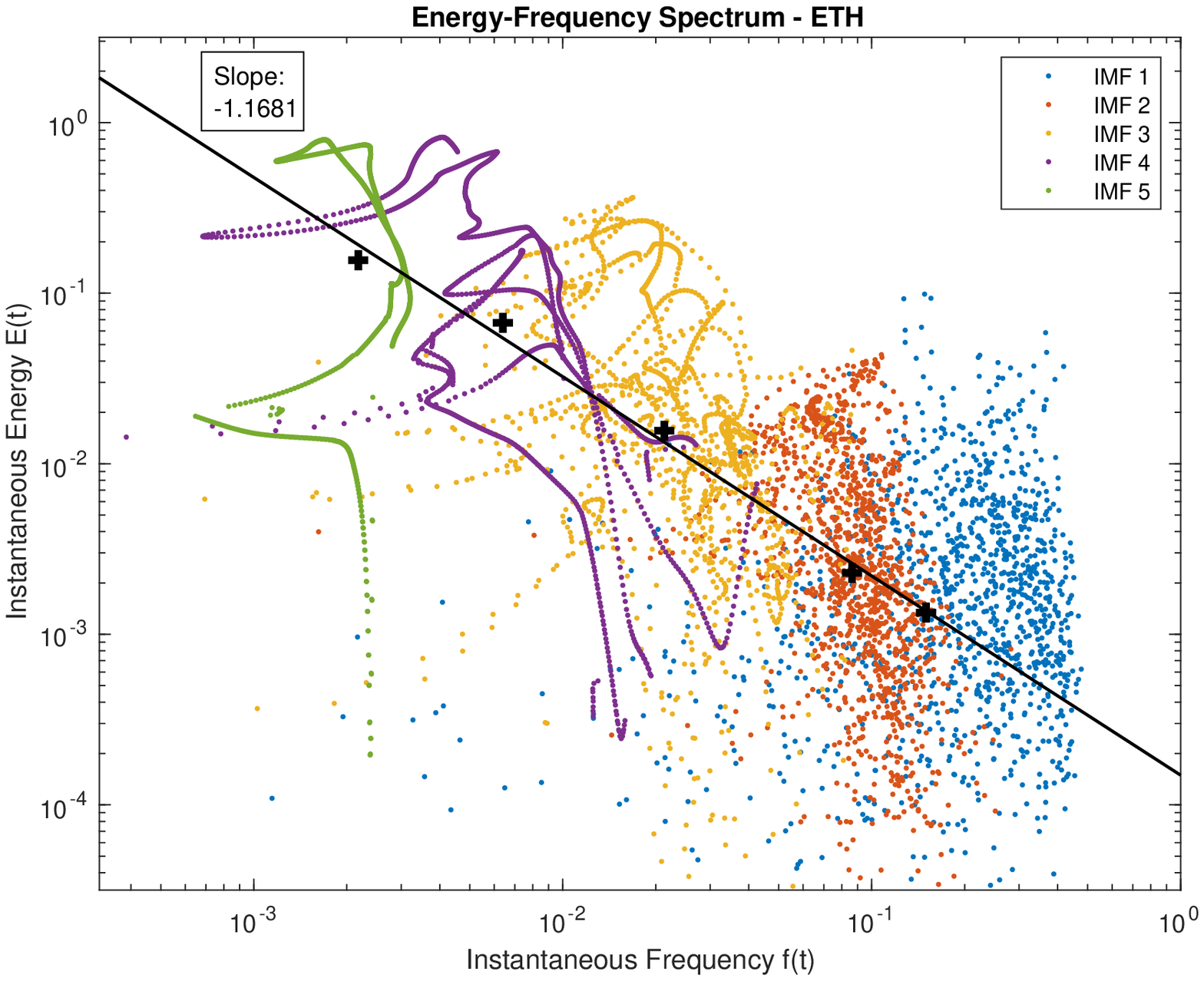} \\

        \includegraphics[trim = 0.5cm   0.5cm   1.3cm  0.5cm, clip,width=3.3in]{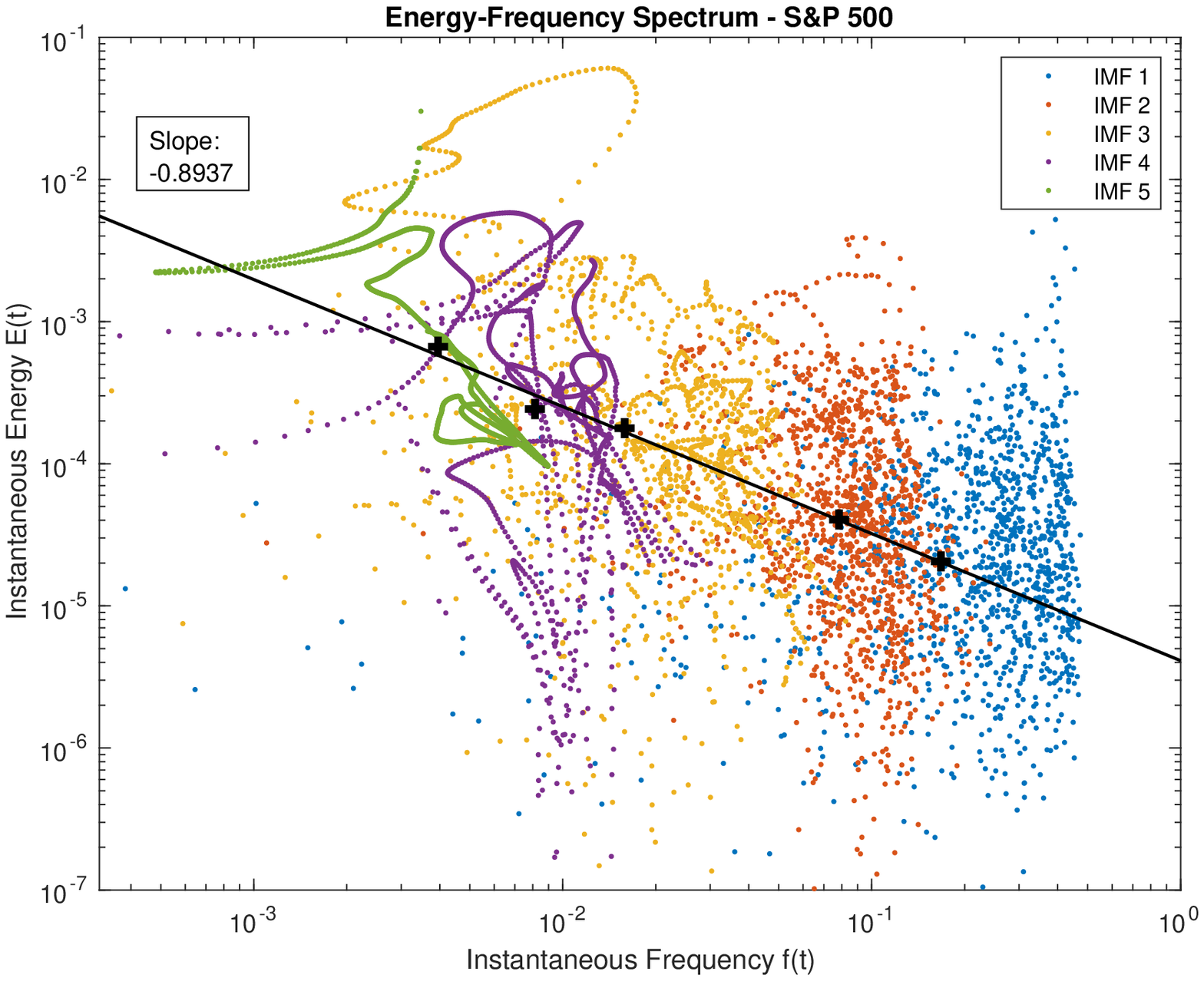}
        \includegraphics[trim = 0.5cm   0.5cm   1.3cm  0.5cm, clip,width=3.3in]{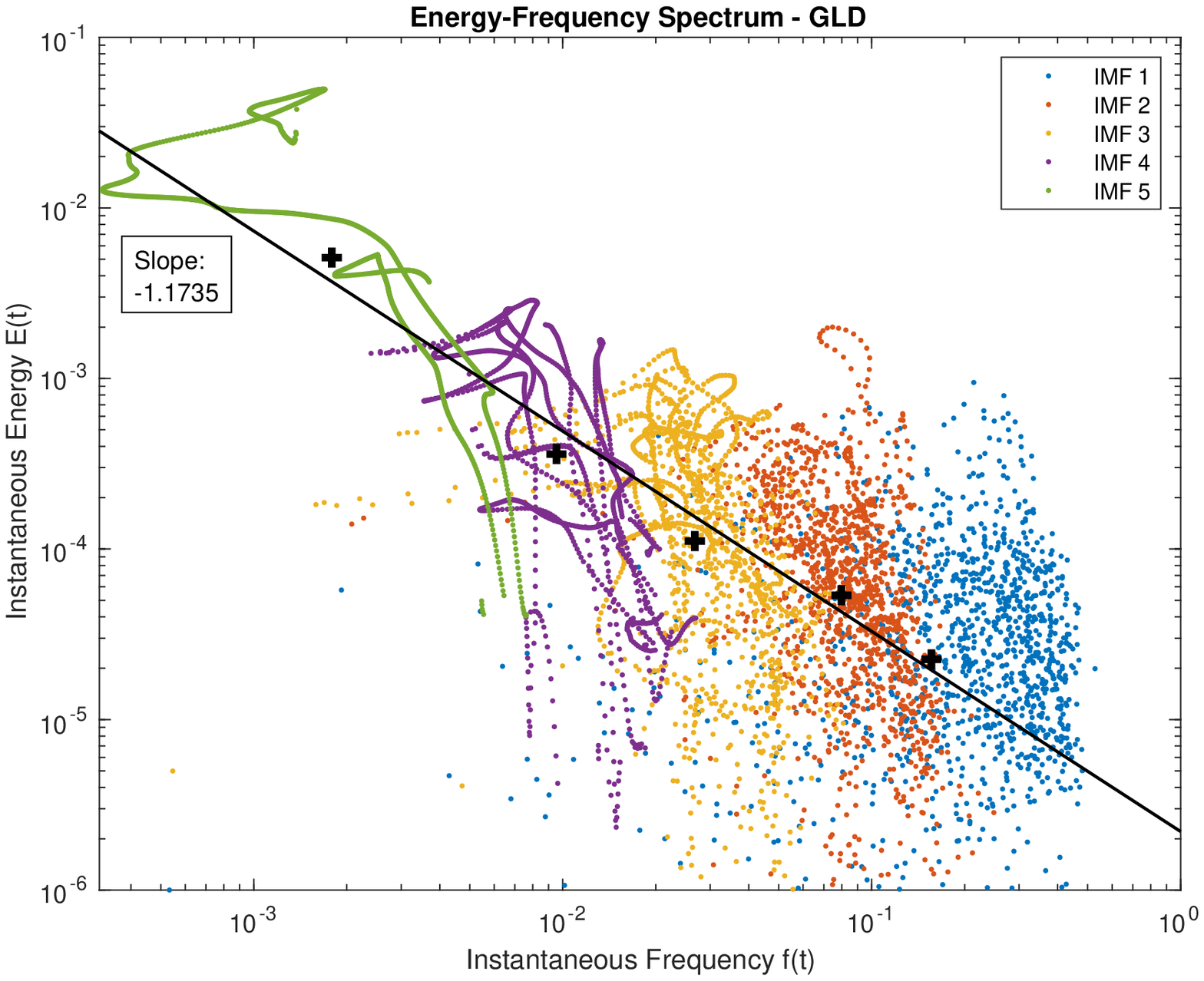}
        \caption{\small{Instantaneous energy-frequency spectrum for two major cryptocurrencies BTC , ETH and two traditional financial market indices S\&P 500 and GLD for comparison. The estimation are done on the time period from 1/4/2016 to 3/29/2021. The black cross in each plot marks the mean of the cluster of points in log space, representing the central frequency for each mode. A linear regression is run on the central frequencies to estimate the slope of the spectrum.}}
    \label{fig:mode_spec}
\end{figure}

For each time series, we can obtain the instantaneous frequency $f_j(t)$ from \eqref{freqEq} and instantaneous energy $E_j(t)$ from \eqref{enrgEq}, corresponding to mode $j=1,\cdots,n$. This allows us to derive the instantaneous  energy-frequency spectrum as shown in Fig. \ref{fig:mode_spec}  for BTC, ETH, LTC, XRP,  S\&P 500, and GLD. Each point on the plots is a pair of $(f_j(t), E_j(t))$, for  mode $j=1,\cdots,n$, and time $t=1,\cdots,T$. We see that for each mode $j$, the instantaneous energy-frequency pairs $(f_j(t),E_{j}(t)), t=1,\cdots,T$ form a cluster of points. 

We define the \emph{central frequency} and \emph{central energy} of mode $j$ during the time period $[0,T]$ as follows:
\begin{align}
\bar{f}_j   &= \exp\left(\frac{1}{T} \int_{0}^{T} \log f_j(t) dt \right), \label{eq_centralFreq}\\
    \bar{E}_j   &= \exp\left(\frac{1}{T} \int_{0}^{T} \log E_j(t) dt \right)\,.\label{eq_centralenergy}
\end{align}
While the instantaneous frequency and energy are time-varying, they are typically fluctuating or orbiting around the central points. \cite{dragomiretskiy2013variational} assume a central frequency in each mode and used the concept to derive the variational mode decomposition (VMD). Intuitively, the central frequency and central energy capture the overall spectral properties of the time series (\cite{wu2004study}).

From Fig. \ref{fig:mode_spec} we also observe a clear linear relationship in the log space of central frequency and energy pair $(\bar{f}_j, \bar{E}_j)$, which are marked as the black crosses. This indicates a power spectrum relation
\begin{equation}
E(f) \sim \frac{1}{f^\alpha}
\end{equation}
The power spectrum exponent $\alpha$ controls how fast the energy decays from lower to higher frequency, and is key to the property of the spectrum and the associated time series. It has been long observed that many time series in nature have $\alpha$ close to one, well-known as the $1/f$ spectrum or ``pink noise." In Fig. \ref{fig:mode_spec} the solid line in each plot is obtained from linear regression of $(\log(\bar{f}_j), \log(\bar{E}_j))$. The negative slope of the line estimates the power spectrum exponent $\alpha$ for the time series. 

Comparing the financial time series, the cryptocurrencies exhibit a more rapid dissipation of energy with respect to frequency than that of S\&P 500 and GLD. This reveals that the rapid fluctuations in the cryptocurrencies tend to have relatively smaller magnitude than those for S\&P 500, which contradicts the notion that the cryptocurrencies are volatile with rapid movements. In terms of power spectrum, S\&P 500 has exponents closest to the $1/f$ spectrum, while the cryptocurrencies and gold ETF generally have $\alpha > 1$. In Table \ref{table:alpha_exp}, we display the $\alpha$ exponents for the top crytocurrencies in terms of market capitalization, and compare against the results from three traditional financial market indices/ETFs. The results show consistently that cryptocurrenties and GLD mostly have $\alpha > 1$, while the equity indices, S\&P 500 and Dow Jones (DJI),  typically have $\alpha$ below and close to 1.

\clearpage
\begin{table}[ht]
\begin{footnotesize}
\centering
\begin{tabular}{c|c|c||c|c|c||c|c|c}
\hline
Crypto &  $\alpha$ & $R^2$ & Crypto &  $\alpha$ & $R^2$ & Crypto &  $\alpha$ & $R^2$ \\
\hline
BTC & 1.2070 & 0.9911 & ETH & 1.1681 & 0.9898 & BNB & 1.2460 & 0.9333 \\
ADA & 1.3294 & 0.9865 & XRP & 1.2307 & 0.9844 & LTC & 1.1832 & 0.9994 \\        THETA & 1.1371 & 0.9591  & LINK & 0.9433 & 0.9200 & BCH & 1.0574 & 0.9887 \\
XLM & 1.1545 & 0.9964 & DOGE & 1.1759 & 0.8701 & TRX & 1.1160 & 0.9621 \\       
MIOTA & 1.0950 & 0.9860 & XMR & 1.0613 & 0.9809 & EOS & 1.0811 & 0.9817 \\
\hline\hline
Index &  $\alpha$ & $R^2$ & Index &  $\alpha$ & $R^2$ & Index &  $\alpha$ & $R^2$ \\
\hline
S\&P 500 & 0.8937 & 0.9924 & DJI & 0.8776 & 0.9191 & GLD & 1.1735 & 0.9669 \\  
\hline
\end{tabular}
\caption{\small{Power spectrum exponents $\alpha$ of top 30 cryptocurrencies by market capitalization, excluding stable coins and coins with less than 3 years of history. At the bottom of the table are equity indices (S\&P 500 and Dow Jones) and gold ETF (GLD) for comparison. Estimations are implemented on the time period 1/4/2016--3/29/2021 (or start from inception dates or earliest available records). }}
\label{table:alpha_exp}
\end{footnotesize}
\end{table}

\subsection{Frequency Synchronization}
\begin{figure}[ht]
    \centering
    \includegraphics[trim = 0.2cm   0cm  0.5cm  0.5cm, clip, width=3.3in]{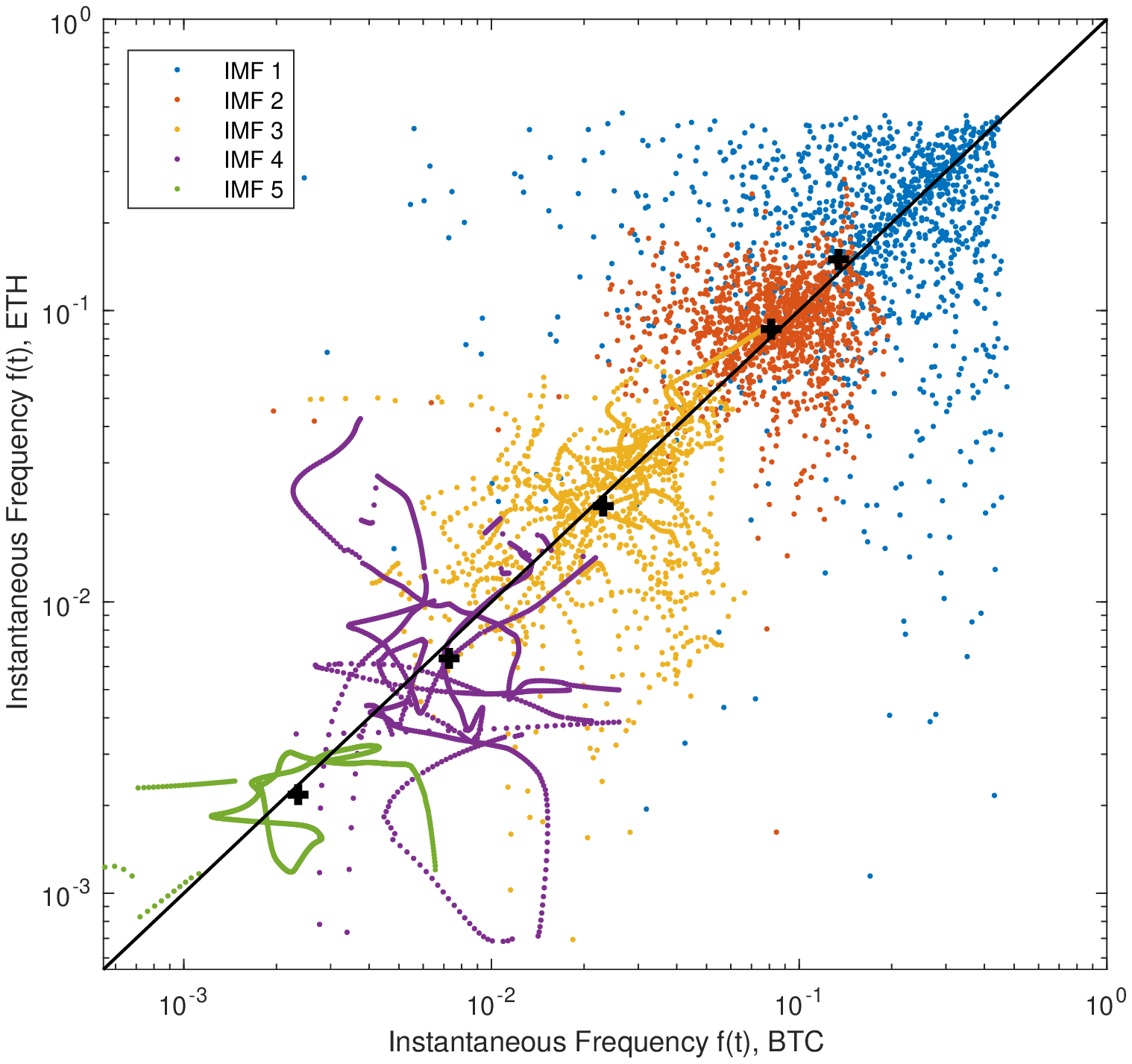}
    \includegraphics[trim = 0.2cm   0cm  0.5cm  0.5cm, clip, width=3.3in]{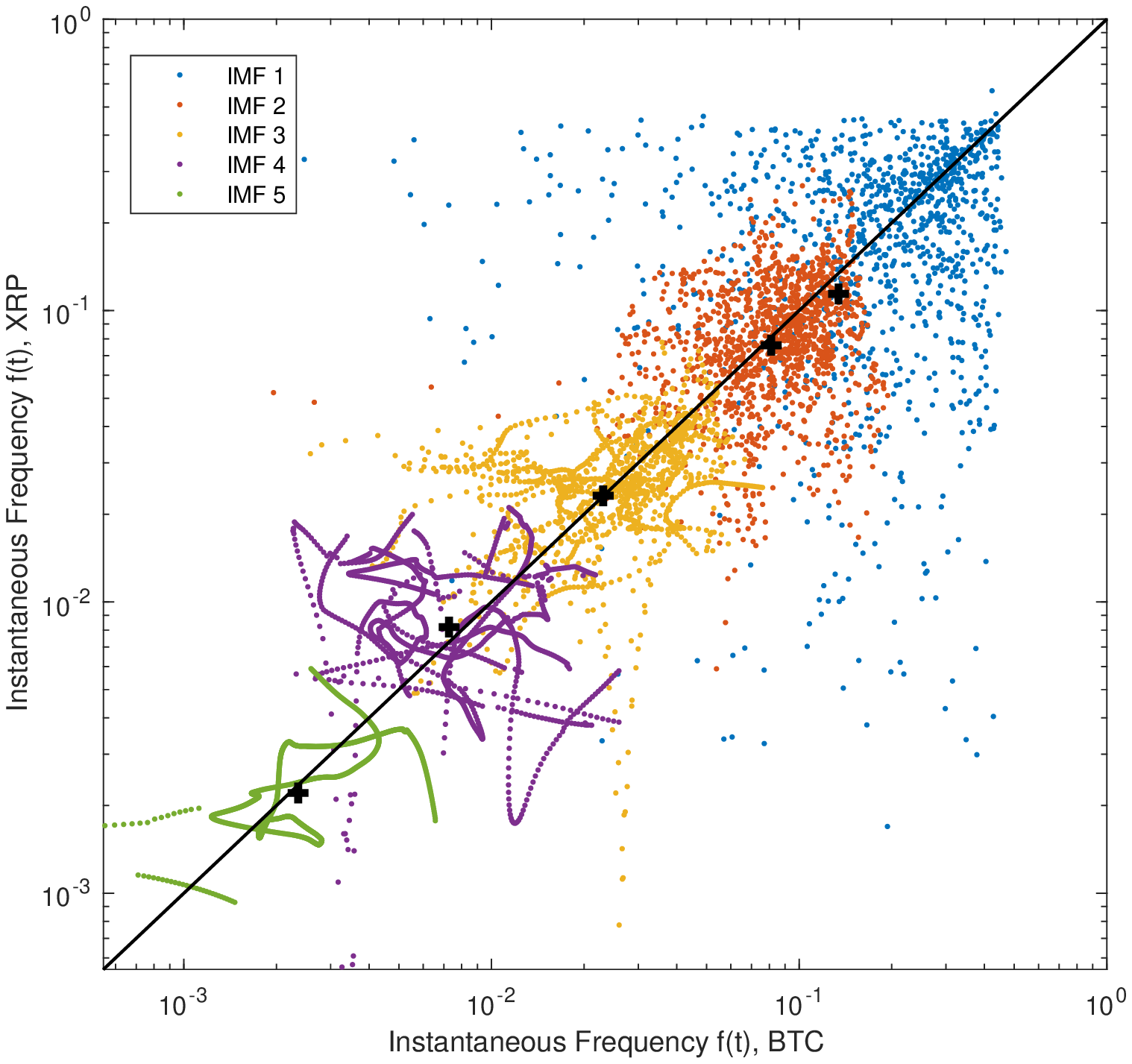} \\
    \includegraphics[trim = 0.2cm   0cm  0.5cm  0.5cm, clip, width=3.3in]{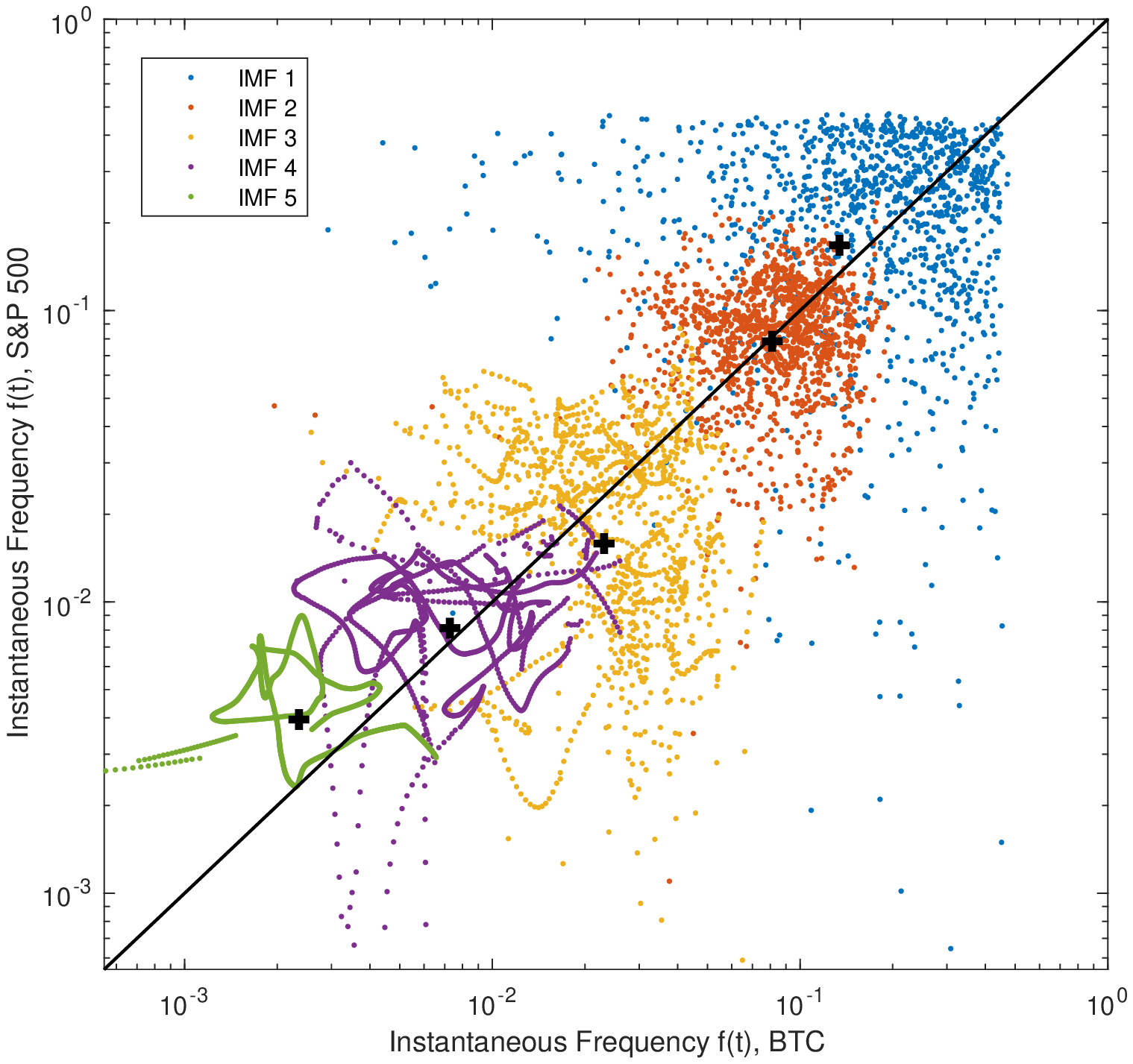}
    \includegraphics[trim = 0.2cm   0cm  0.5cm  0.5cm, clip, width=3.3in]{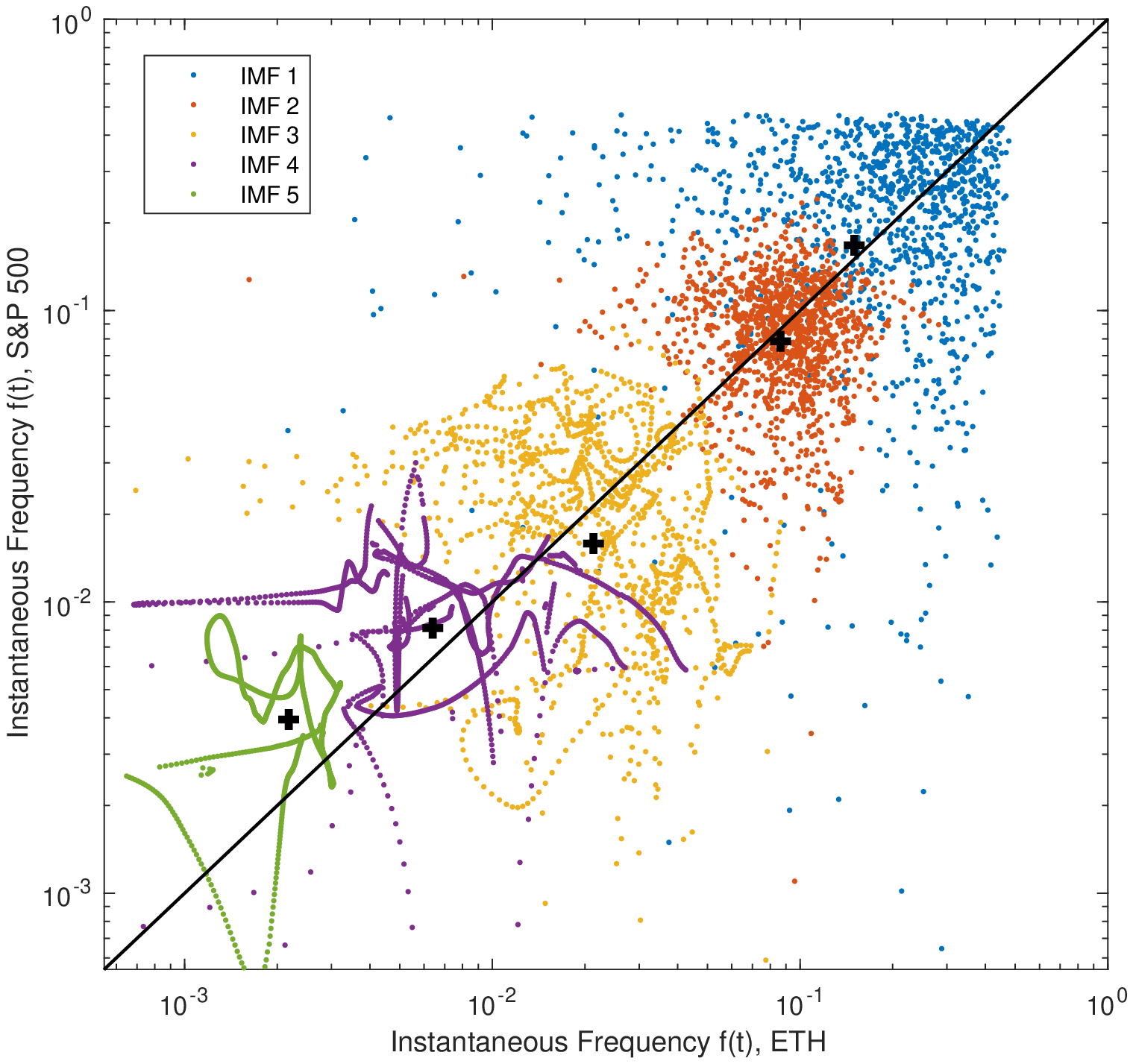}
        \caption{{Instantaneous frequency: BTC v.s. ETH (top left), BTC v.s. XRP (top right), BTC v.s. S\&P 500 (bottom left), and ETH v.s. S\&P 500 (bottom right). The estimations are done on the time period from 1/4/2016 to 3/29/2021.}}
    \label{fig:freq_freq}
\end{figure}

To further investigate the spectral difference between cryptocurrencies and  traditional financial market, we compare between each two time series the central frequencies $\bar{f}_j, \ j=1,\cdots,n$. In Fig. \ref{fig:freq_freq},  we show the log-log scatter plots of the instantaneous frequencies of four asset pairs: BTC vs ETH, BTC vs XRP, BTC vs S\&P 500, and ETH vs S\&P 500. Each point on the plot is a pair of instantaneous frequencies of the two time series recorded at the same time, and the central frequencies are marked as the black crosses. The solid straight line of unit slope shows the reference line for identical frequencies. We observe that the cryptocurrencies, e.g. BTC vs ETH and BTC vs XRP, share very similar frequency profiles. The central frequencies of all the IMF components are very close between two cryptocurrencies. On the other hand, the central frequencies deviate from the identical line at low frequency modes for BTC vs S\&P 500 and ETH vs S\&P 500. More specifically, the cryptocurrencies show slower dynamics in the longer term components, while the fast modes of both cryptocurrencies and S\&P 500 have very similar mean frequencies. 


The central frequencies of the IMF components show consistent similarity within the cryptocurrency market and common deviation from the traditional stock market. The interesting identical frequency profile within the cryptocurrency market suggests synchronization, which is a typical phenomenon in nonlinear dynamics with interaction (see e.g. \cite{pikovsky2003synchronization}). In order to quantify  the frequency synchronization level between two  time series $x_1$ and $x_2$, we  define the  associated \emph{frequency deviation}  as follows:

\begin{equation}\label{eq-dev}
D(x_1,x_2) := \sum_{j=1}^n \log^2\left(\frac{\bar{f}^{(1)}_j}{\bar{f}^{(2)}_j}\right)  = \|\log(\bar{f}^{(1)}) - \log(\bar{f}^{(2)})\|^2.
\end{equation}
A lower frequency deviation value represents higher synchronization level. In particular, we have $D(x_1,x_2) = 0$ if and only if the central frequencies of all the IMF components are identical for $x_1$ and $x_2$, meaning the two time series are fully synchronized.

\clearpage

\begin{figure} [ht]
 \centering
    \includegraphics[width=6.5in]{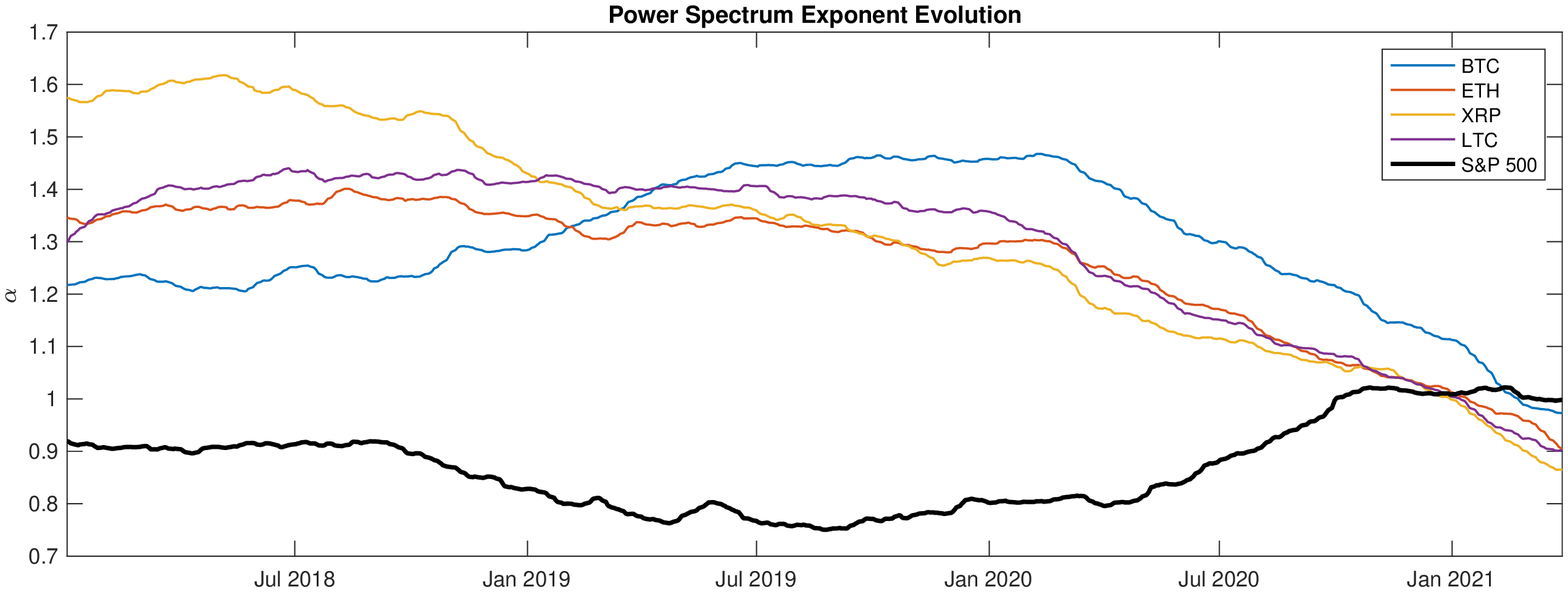} \\
    \includegraphics[width=6.5in]{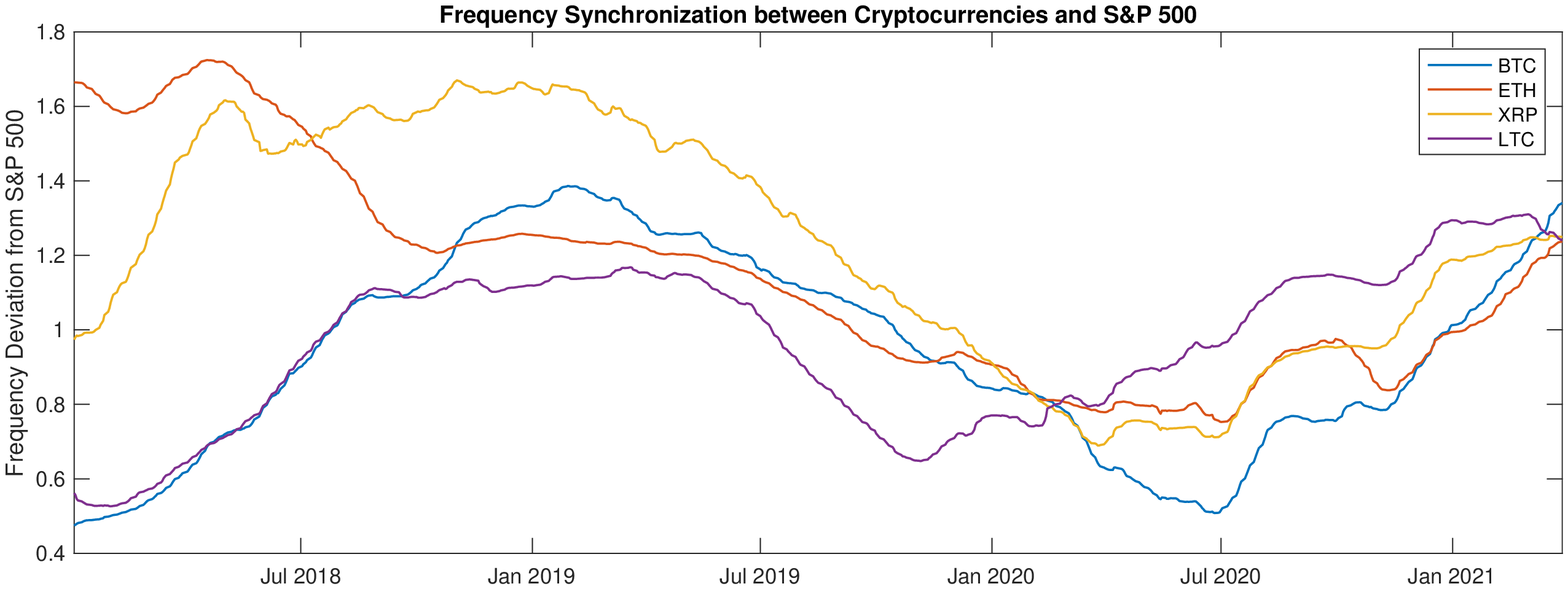}
        \caption{\small{Top: The evolution of the power spectrum exponents for four major cryptocurrencies  $\{$BTC, ETH, XRP, LTX$\}$, compared against S\&P 500. Bottom: The frequency deviations from S\&P 500 for the 4 cryptocurrencies. In both plots, results are computed based on 2-year rolling windows from 1/4/2016 to 3/29/2021.}}
    \label{fig:alpha_evo}
\end{figure}

   It is of interest to see how the cryptocurrencies have been evolving compared against the stock market represented by S\&P 500. In Fig. \ref{fig:alpha_evo}, we plot the estimated power spectrum exponent $\alpha$ based on a 2-year rolling window, for the cryptocurrencies BTC, ETH, LTC, XRP, and the benchmark S\&P 500. We see that the power spectrum exponents $\alpha$  of cryptocurrencies are moving closer to that of S\&P 500. Furthermore, in the second plot, we see that the frequency deviation from S\&P 500 defined in \eqref{eq-dev} has decreased from July 2017 till July 2020 for all the four cryptocurrencies. These observations are evidence of synchronization over a relatively long period of time and suggest an underlying interaction between cryptocurrencies and the stock market.\\

\section{Conclusion}\label{sect-conclude}
 We have presented the ACE-EMD method for multiscale analysis of  nonstationary financial time series with a focus on cryptocurrency prices. The key outputs of this method are the series of intrinsic mode functions, along with the time-varying instantaneous amplitudes and instantaneous frequencies. Different combinations of modes allow us to reconstruct the financial time series based on different timescales. This allows us to better understand cryptocurrency price movements due to short-term fluctuations vs. long-term trends. It also sheds light on the multiscale properties of cryptocurrency volatilities that set them apart from traditional equities. In the same spirit, we apply Hilbert spectral analysis to  compute  the associated instantaneous energy-frequency spectrum for a collection of cryptocurrencies.   In particular, the  power spectrum exponents of cryptocurrencies have been significantly higher than that of  S\&P 500, but are seen to be converging in recent months.

For future research, one practical application of the ACE-EMD method is to generate a collection of new features that can be integrated into machine learning models. Several studies have utilized HHT together with machine learning for time series forecasting (\cite{kurbatsky2014,LeungZhao_HHT}).
This approach using CEEMD can be very useful not only for financial forecasting but also for cryptocurrency portfolio construction (\cite{leung_nguyen}).

\bibliographystyle{apa}
\begin{small}
\bibliography{mybib_NEW}
\end{small}
\end{document}